\documentclass[iop,letterpaper]{emulateapj}
\newcommand{\dfplot}[1]{\plotone{figs/#1}}
\newcommand{\dfplottwo}[2]{\plottwo{figs/#1}{figs/#2}}
\newcommand{\dfplots}[1]{\begin{center}\includegraphics{figs/#1}\end{center}}

\newcommand{\IRAS}{{\it IRAS}}

\newcommand{\HI}{H\,{\scriptsize I}}

\newcommand{\MgII}{Mg$_2$ }

\newcommand{\etal}{{\it et al.}~}

\newcommand{\erf}{{\rm erf}}

\newcommand{\degree}{\ensuremath{^\circ}}

\newcommand{\MAG}{{\rm ~mag}}
\newcommand{\MMAG}{{\rm ~mmag}}

\newcommand{\pc}{{\rm ~pc}}
\newcommand{\EBVSFD}{E(B-V)_{\mathrm{SFD}}}
\newcommand{\Xsfd}{\mathbf{X}_{\mathrm{SFD}}}
\newcommand{\Xfiras}{\mathbf{X}_{\mathrm{FIRAS}}}
\newcommand{\mum}{~\ensuremath{\mu \mathrm{m}}}
\newcommand{\Aeff}[1]{A(\lambda_{\mathrm{eff},#1,S})}
\newcommand{\tauum}[1]{\tau_{#1\mum}}

\begin{document}

\title{The Blue Tip of the Stellar Locus:\\
       Measuring Reddening with the SDSS}

\author{Edward F. Schlafly\altaffilmark{1}, 
Douglas P. Finkbeiner\altaffilmark{2},
David J. Schlegel\altaffilmark{3},
Mario Juri\'{c}\altaffilmark{2,4},
\v{Z}eljko Ivezi\'{c}\altaffilmark{5},
Robert R. Gibson\altaffilmark{5},
Gillian R. Knapp\altaffilmark{6},
Benjamin A. Weaver\altaffilmark{7}
}

\altaffiltext{1}{Department of Physics, Harvard University, 17 Oxford Street, Cambridge, MA 02138, USA}
\altaffiltext{2}{Harvard-Smithsonian Center for Astrophysics, 60 Garden Street, Cambridge, MA 02138, USA}
\altaffiltext{3}{Lawrence Berkeley National Lab, 1 Cyclotron Road, MS 50R5032, Berkeley, CA 94720, USA}
\altaffiltext{4}{Hubble Fellow}
\altaffiltext{5}{Department of Astronomy, University of Washington, Box 351580, Seattle, WA 98195, USA}
\altaffiltext{6}{Department of Astrophysical Sciences, Princeton University, Peyton Hall, Princeton, NJ 08544, USA}
\altaffiltext{7}{Department of Physics, New York University, 4 Washington Place, New York, New York 10003, USA}

\begin{abstract}
We present measurements of reddening due to dust using the colors of stars in the Sloan Digital Sky Survey (SDSS).  We measure the color of main sequence turn-off stars by finding the ``blue tip'' of the stellar locus: the prominent blue edge in the distribution of stellar colors.  The method is sensitive to color changes of order 18, 12, 7, and 8$\MMAG$ of reddening in the colors $u-g$, $g-r$, $r-i$, and $i-z$, respectively, in regions measuring $90^\prime$ by $14^\prime$.  We present maps of the blue tip colors in each of these bands over the entire SDSS footprint, including the new dusty southern Galactic cap data provided by the SDSS-III.  The results disfavor the best fit \citet{O'Donnell:1994} and \citet{Cardelli:1989} reddening laws, but are well described by a \citet{Fitzpatrick:1999} reddening law with $R_V = 3.1$.  The \citet[SFD]{Schlegel:1998} dust map is found to trace the dust well, but overestimates reddening by factors of 1.4, 1.0, 1.2, and 1.4 in $u-g$, $g-r$, $r-i$, and $i-z$, largely due to the adopted reddening law.  In select dusty regions of the sky, we find evidence for problems in the SFD temperature correction.  A dust map normalization difference of 15\% between the Galactic north and south sky may be due to these dust temperature errors.

\emph{Subject headings: }
dust, extinction --- 
ISM: clouds  --- 
Galaxy: stellar content
\end{abstract}
\maketitle

\section{Introduction}
Most astronomical observations are affected by Galactic interstellar dust, whether as a source of foreground light in the microwave, far-infrared and gamma-ray wavelength regions or as a cause of extinction in the infrared through ultraviolet \citep{Draine:2003}.  Characterizing the properties of the dust and accounting for its effects on observations is then a central problem in astronomy.

Dust is formed as stars burn nuclear fuel to heavy elements and emit these elements in stellar winds or in more violent eruptions, and these elements are reprocessed in the interstellar medium \citep{Draine:2009}.  The distribution of dust is correspondingly correlated with the hydrogen in the interstellar medium (ISM), though the dust-to-gas ratio varies.  \citet[BH]{Burstein:1978} took advantage of the correlation between \HI~and dust column density to make the first widely used map of dust column density, combining \HI~emission and galaxy counts.

The BH dust map was superseded by the \citet[SFD]{Schlegel:1998} dust map, which took advantage of the full-sky far-infrared (FIR) data provided by {\it IRAS} and DIRBE, which are dominated by thermal emission from the Galactic dust at wavelengths of 100$\mum$ and longer.  After correction for dust temperature using FIR color ratios, these maps trace dust column density.  Calibration of dust column density to color excess $E(B-V)$ was performed using the colors of a sample of 389 galaxies with \MgII indices \citep{Faber:1989}, which were also used to test the performance of the resulting map.

The advent of large astronomical surveys permits stronger tests of the SFD dust map and the assumptions used to construct it.  We test the SFD dust map using the colors of stars from the Sloan Digital Sky Survey (SDSS) \citep{York:2000}, including the recently completed SDSS-III imaging \citep{Weinberg:2007}, which covers about 2000 deg$^2$ in the southern Galactic sky.  This allows us to more tightly constrain the SFD normalization and the dust extinction spectrum, or ``reddening law,'' over the SDSS bands, as well as to provide a map of color residuals that point to problems with the SFD dust map, and particularly with the temperature correction.

Previous tests of SFD have usually found that SFD overpredicts extinction in high-extinction regions.  Shortly after its introduction, \citet{Arce:1999} found that SFD extinction was too high by 40\% in Taurus, using star counts, colors, and FIR emission.  Likewise, studies using globular cluster photometry, galaxy counts, and NIR colors have found that SFD overpredicts extinction by a similar fraction in other dusty regions \citep{Stanek:1998, Chen:1999, Yasuda:2007, Rowles:2009}.  \citet{Cambresy:2001} explore the link between the SFD overestimation and dust temperature using star counts in the Polaris Flare.  At $|b| < 40\degree$, \citet{Dobashi:2005} use optical star counts to conclude that SFD overpredicts extinction by a factor of two or more.  We perform similar tests to these in regions of lower extinction than had been previously possible, taking advantage of the high quality and depth of the SDSS stellar photometry, and complementing the recent work of \citet{Peek:2010}, who use SDSS galaxy spectra.  In these regions, we do not find that SFD overpredicts reddening by a large factor; rather, we find that SFD overpredicts reddening by about 14\% in $B-V$, though, because of the reddening law adopted by SFD, this varies from color to color.

The colors of stars vary substantially with stellar type, and with location in the Galaxy due to the effect of metallicity on color.  Nevertheless, we find that the colors of the most blue main sequence stars in old populations---the main sequence turn-off (MSTO) stars---are remarkably stable over the sky, and that we can empirically model their slow spatial variation.  We therefore present measurements of the colors of the ``blue tip'' of the stellar locus, and use them to constrain the SFD dust map.  This work is akin to that of \citet{High:2009}, in which the color of stellar populations is also used as a sort of color standard.

In \textsection \ref{sec:data} we describe the data sets used in this work: the SDSS imaging data and the SFD dust map.  In \textsection \ref{sec:bluetip}, we present our method for measuring the blue tip of the stellar locus, and the corresponding maps of blue tip colors on the sky.  In \textsection \ref{sec:fittingthemap}, we present fits of the SFD dust map to the blue tip colors.  In \textsection \ref{sec:discussion} and \textsection \ref{sec:conclusion}, we discuss these results and conclude.  The blue tip maps and measurements can be found at the web site \texttt{skymaps.info/bluetip}.

\section{Data}
\label{sec:data}

\subsection{The SDSS}

The SDSS is a digital spectroscopic and photometric survey, that, with the additional south Galactic cap (SGC) data provided by the SDSS-III, covers just over one third of the sky, mostly at high latitudes \citep{Abazajian:2009}.  The SDSS provides near-simultaneous imaging in five optical filters: $u$, $g$, $r$, $i$, and $z$ \citep{Gunn:1998, Fukugita:1996}.  The photometric pipeline has uniformly reduced data for about $10^8$ stars.  The SDSS is 95\% complete up to magnitudes 22.1, 22.4, 22.1, 21.2, and 20.3 in $u$, $g$, $r$, $i$, and $z$.  The SDSS imaging is performed in a drift-scanning mode in which the shutter is left open while the telescope moves at constant angular speed across the sky. The resulting imaging ``runs'' tend to cover strips of sky that are long in right ascension and narrow in declination.  These runs are divided into fields, which are $13.5^\prime$ by $9^\prime$ in size.  We use SDSS data that have been photometrically calibrated according to the ``ubercalibration'' procedure of \citet{Padmanabhan:2008}.

\subsection{The SFD Dust Map}
The SFD dust map is a map of thermal emission from dust based on the \IRAS\ 100$\mum$ maps.  The \IRAS\ 100$\mum$ map underwent three major processing steps before being used in the SFD dust map: it was destriped, zodiacal-light subtracted, and calibrated to match COBE/DIRBE at degree scales.  The first and third of these steps were necessary because the zero point of the \IRAS\ 100$\mum$ detector varied over time, imprinting the \IRAS\ scan pattern on the data in the form of stripes, and making the overall zero point of the data uncertain.  The zodiacal-light subtraction is needed to remove the signature of the hot, bright, local interplanetary dust from the \IRAS\ and DIRBE data, which, though bright, contributes negligibly to the reddening.  The DIRBE 100$\mum$ to 240$\mum$ flux ratio is used to constrain the dust temperature and to transform the 100$\mum$ flux map to a map proportional to dust column density.  The destriped \IRAS\ data have resolution of 6 arcminutes, while the temperature correction is smoothed to degree scales.  The destriped, temperature-corrected \IRAS\ 100$\mum$ map is normalized to a map of $E(B-V)$ by using a sample of 389 galaxies with \MgII indices and $B-V$ colors, for use as color standards \citep{Faber:1989}.

\section{The Blue Tip of the Stellar Locus}
\label{sec:bluetip}

\begin{figure*}[tbh]
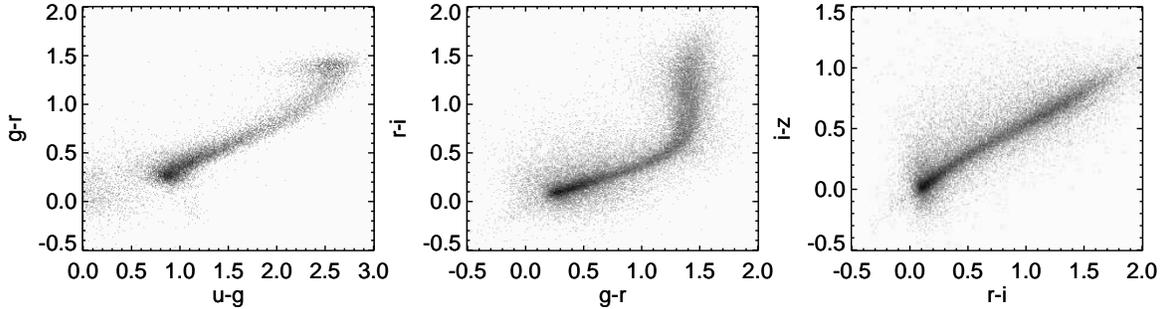

\dfplot{colorplotsccd.ps}
\figcaption{
\label{fig_ccd}
Color-color diagrams for point sources from the SDSS, within two degrees of the north Galactic pole.  The sharp cutoff in stellar density blueward of the ``blue tip'' is prominent.  In $u-g$, the points blueward of the blue tip are mostly quasars and white dwarfs. The blue tip is at about 0.8, 0.2, 0.1, and 0.0 mags in $u-g$, $g-r$, $r-i$, and $i-z$, respectively.
}
\end{figure*}

Main sequence stars are confined to a tight locus in color-color space, as is apparent from a typical SDSS color-color diagram (Fig.~\ref{fig_ccd}).  The locus has a sharp blue edge at the color of the MSTO stars, blueward of which there are only rare blue stragglers, white dwarfs, and quasars.  The intrinsic color of the MSTO is set by the properties of the stellar population: primarily, metallicity and age.  The observed colors of the MSTO are the intrinsic colors, altered by reddening due to dust and by systematic color shifts due to imperfect calibration of the observations.  Empirically, the intrinsic color of the MSTO is slowly varying in space; the observed color of MSTO stars is therefore a good probe of the reddening due to dust.

\begin{figure*}[tbh]
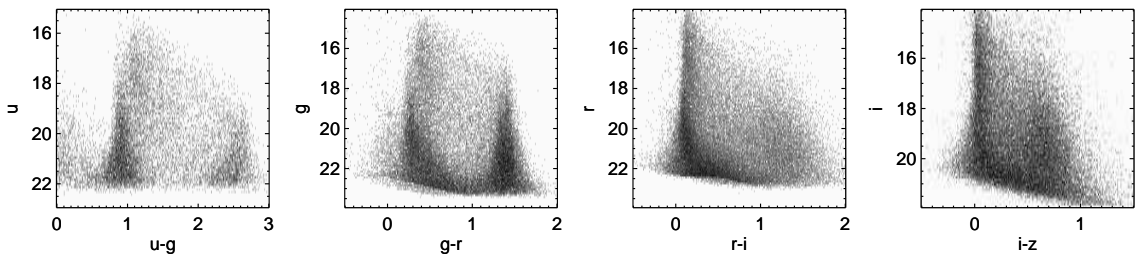

\dfplot{colorplotscmd.ps}
\figcaption{
\label{fig_cmd}
Color-magnitude diagrams from the SDSS, for the same stars as in Fig.~\ref{fig_ccd}.  The location of the blue edge of the stellar locus depends on magnitude, especially in the bluer bands, where bright, nearby, more metal-rich stars are redder than more distant halo stars.  In $u-g$, blueward of the blue edge lie mostly quasars and white dwarfs.  The location of the blue tip is as in Fig.~\ref{fig_ccd}.
}
\end{figure*}

Population effects cause the color of main sequence stars to vary as a function of the magnitude of stars probed.  At faint magnitudes and high latitudes, metal-poor halo stars dominate, rendering the MSTO stars bluer than at bright magnitudes and lower latitudes, where redder disk stars dominate (Fig.~\ref{fig_cmd}).  Moreover, the shape of the blue edge changes with magnitude: owing to photometric errors, the edge is sharp for bright stars and becomes blurred at fainter magnitudes.  Accordingly, the typical color of MSTO stars depends on magnitude.

Stars of similar metallicities may clump together in the halo, changing the observed color of the blue tip.  Because we do not have a good way to model such effects, we cannot distinguish such clumps from reddening due to dust in the blue tip maps.

\subsection{Measuring the Blue Tip}

We perform fits to the location of the blue tip of the stellar locus in each SDSS field (or group of SDSS fields).  To avoid biasing the fits, we select stars in each field for analysis in a reddening-independent way, and then find the location of the blue tip in the resulting sample.

Care must be taken when imposing cuts on the stars in each field to avoid biasing the resulting blue tip color.  Because extinction changes the observed magnitudes of stars, a flat cut on magnitude would bias the measurement, as the distance range probed would change with the dust column.  Instead, we perform cuts on $D$, the $g$ magnitude of a star projected along the reddening vector to zero color:
\begin{equation}
D = g - \frac{A_g}{A_g-A_r}(g-r).
\end{equation}
Provided $A_g/(A_g-A_r)$ is accurate, the set of stars in a particular range of $D$ is independent of the dust column towards the stars.  $D$ is related to the distance modulus of the MSTO stars (though not redder stars).  Using the \citet{Juric:2008} ``bright'' photometric parallax relation, MSTO stars with $r-i = 0.1$ and $g-r = 0.2$ have $M_g = 5.3$.  Accounting for the projection from $g-r = 0.2$ to zero color along the reddening vector, $D \approx g-M_g+4.6$.  In this work, we frequently use the range $10 < D < 19$.  Including the effect of saturation on the bright end, this restriction on $D$ selects stars with distances between about 1 and 8 kpc.

\begin{figure*}[tbh]
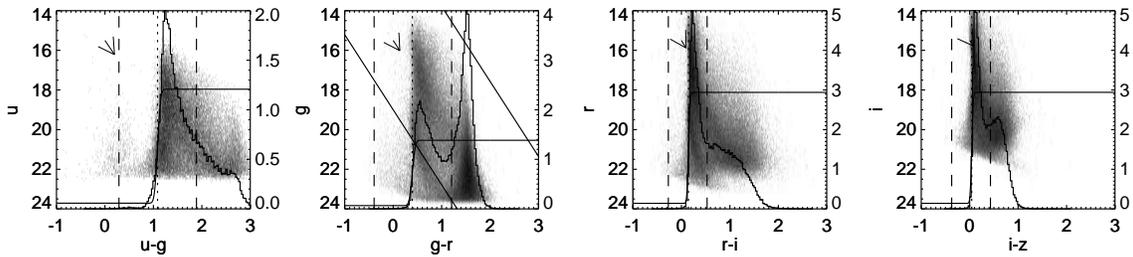

\dfplot{blueplots.ps}
\figcaption{
\label{fig_blueplot}
Example fits for the blue tip of the stellar locus in $u-g$, $g-r$, $r-i$, and $i-z$ for a one degree region around $l = 150\degree$ and $b = 20\degree$.  A grayscale of the color-magnitude diagram is plotted, with features of the fit superimposed.  The diagonal solid lines in the $g-r$ plot give the magnitude range within which stars are fitted, corresponding to $10 < D < 19$.  The histograms give the distribution of stellar colors satisfying this cut.  The solid error function gives the best fit to this histogram, within the gray dashed vertical lines.  The black dotted vertical line gives the derived location of the blue tip.  The arrow in the upper left of the diagram gives the magnitude and direction of the reddening vector in this field, according to SFD.  Finally, the right hand axis gives the value of the pdf $P$; its integral between the outer dashed lines is unity.
}
\end{figure*}

With a reddening-independent population of stars in hand, we measure the color of the blue edge of these stars in each SDSS color (Fig.~\ref{fig_blueplot}).  The stellar colors are modeled as drawn from a probability distribution $P(x)$, taken to be a step function convolved with a Gaussian to reflect the photometric uncertainty and the intrinsic width of the blue edge.  The maximum-likelihood location for the step is denoted the ``blue tip'' of the locus.  Specifically, the form of the probability distribution is taken to be
\begin{equation}
P= \frac{1}{2}\left[1+\erf\left(\frac{x-x_0}{\sqrt 2 \sigma_P} \right)\right] + F,
\end{equation}
where erf is the Gaussian error function, $\sigma_P$ corresponds to the width of the edge and $F$ sets a floor to the probability distribution, to render the fit insensitive to the occasional white dwarf or blue straggler.  In this work we use $F=0.05$.  The variable $x_0$ is the only free parameter in the fits, and gives the color of the blue tip.  The final results of this work are insensitive to substantial changes in the floor $F$, from 0.01 to 0.05.  The formal statistical uncertainty in the blue tip color is computed for each measurement by fitting the likelihood near the maximum likelihood to a Gaussian.

The width of the blue edge is set by a combination of the intrinsic width of the edge and the typical measurement error in star colors in the SDSS.  We adopt the values 0.05, 0.05, 0.025 and 0.025 for the width parameter $\sigma_P$ in the colors $u-g$, $g-r$, $r-i$ and $i-z$, respectively.  Because the photometric uncertainty becomes larger at faint magnitudes, at high extinctions the observed blue edge is expected to broaden.  Despite this, $\sigma_P$ is kept constant in the fits, introducing the possibility of bias in the measurements.  Nevertheless, as we look only at relatively bright ranges of $D$ and low extinction, this is a minor effect.  Simulations with a mock star catalog (\textsection \ref{subsec:mockcatalog}) verify that for the range $10 < D < 19$ used in this work, the introduced bias in the recovered ratio of blue tip color to $E(B-V)$ is less than 2\% in $u-g$ and smaller in the other colors.

The statistical uncertainty of the fit depends on the number of stars included in the fit, and so on the local stellar density, the sky area and range of $D$ used.  At high latitudes, binning together ten SDSS fields, for stars with $10 < D < 19 \MAG$, typical statistical uncertainties are 20, 15, 6 and 6$\MMAG$ for $u-g$, $g-r$, $r-i$, and $i-z$, respectively.  Finding the standard deviations of measurements taken near the north Galactic pole, where the expected reddening is smooth and small, we find uncertainties of 24, 17, 10, and 12$\MMAG$, similar to the fit results.

\begin{figure*}[tbh]
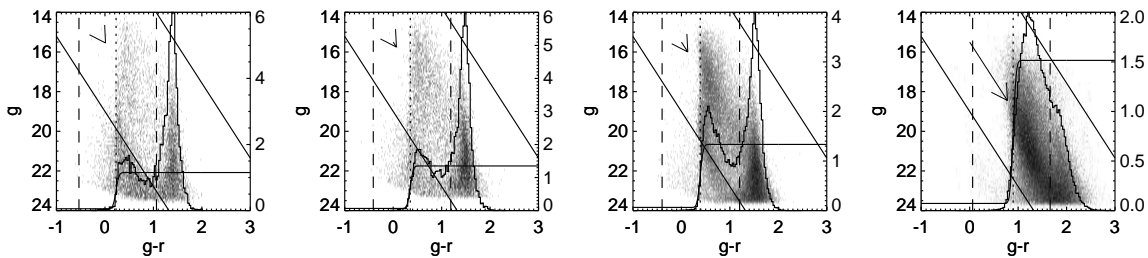

\dfplot{blueplots-ext.ps}
\figcaption{
\label{fig_blueplot-ext}
Example fits for the blue tip of the stellar locus in regions of increasing extinction, at $l = 150\degree$ and $b = 80\degree, 40\degree, 20\degree$ and $10\degree$, from left to right.  The SFD extinction prediction is indicated by the length of the arrow in the upper left of the diagram.  As we approach the plane, the stellar density increases.  At $b=10\degree$, the brightest stars are not behind all of the dust, and so the color of the blue tip becomes redder at fainter magnitudes.
}
\end{figure*}

The presence of dust reddens stars, pushing the blue edge redward (Fig.~\ref{fig_blueplot-ext}).  In low-extinction regions where all of the dust is closer than about 1 kpc, the observed blue tip is uniformly shifted.  It is this shift that we track in our measurements.  In high-extinction regions or regions with dust beyond about 1 kpc, we can in principle learn about the change in dust column with distance by varying $D$, though we have not done so in this work (see \textsection \ref{subsec:dustdistance}). 

The functional form $P(x,x_0)$ chosen to fit the blue edge of the distribution is usually a close fit only near the blue edge itself, and is rather bad over most of the color range.  However, experiments with other functional forms and attempts to adaptively shrink the range over which the fit is performed to a narrow region around the edge tend to render the fit less robust.  Moreover, because $P$ is approximately constant away from the edge, the fit results are robust to outliers far from the edge.  Alternative functional forms systematically shift the location of the edge around slightly (mmags), but not in a way that correlates with the dust.

The blue tip fit can occasionally fail to find the blue tip of the stellar locus when too few stars are used.  In such cases it might identify a single blue star or quasar as the blue tip, or latch on to the M-dwarf peak in color magnitude diagrams.  In the maps we present, with $10 < D < 19$ and using 10 SDSS fields worth of stars for each blue tip fit, fewer than 1\% of measurements are affected.

\subsection{Changing Extinction with Distance}
\label{subsec:dustdistance}

In this work we treat all MSTO stars observed by the SDSS with $10 < D < 19$ as behind all of the dust.  In principle, however, by comparing the color of the blue tip for nearby stars and distant stars, distant clouds of dust could be detected.  Such clouds certainly exist, as dust has been detected in the halos of other galaxies and in \HI\ clouds outside the disk in our Galaxy \citep{Menard:2010, Wakker:1996}.  However, comparing blue tip maps made for nearby ranges of $D$ and distant ranges of $D$ reveal no readily identifiable structures larger than the noise in the halo.  We note that if such structures vary slowly spatially, then they will be difficult to distinguish from variation in blue tip colors due to changing stellar populations.  We defer to later work the attempt to identify such clouds.

On most sight lines the great majority of the dust column comes from the Galactic disk and is associated with the \HI\ disk, which has a scale height of about $150\pc$ \citep{Kalberla:2009}.  Accordingly, as the brightest unsaturated MSTO stars are approximately a kiloparsec away, at high Galactic latitudes even the nearest MSTO stars observed by the SDSS are behind this dust.

\begin{figure*}[tbh]
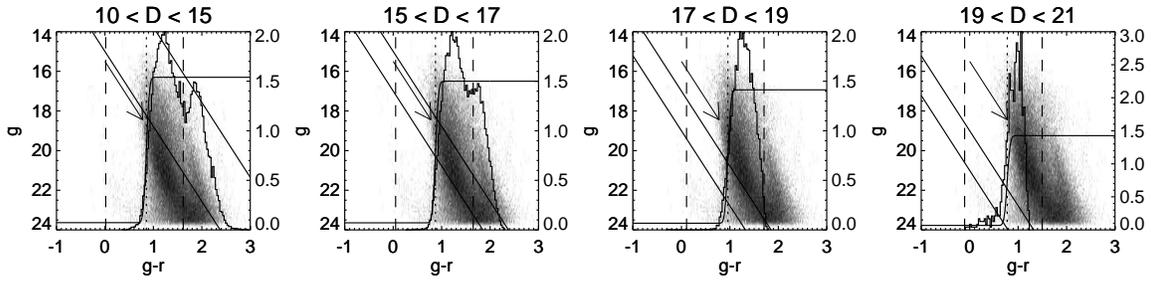

\dfplot{blueplots-dist.ps}
\figcaption{
\label{fig_blueplot-dist}
Example fits for the blue tip of the stellar locus for different magnitude ranges $D$ at $l=150\degree$ and $b=10\degree$.  The blue tip shifts redward as the range of $D$ probed goes fainter, until we run out of stars.  Additionally, intrinsically red foreground stars increasingly contaminate the blue tip as we go fainter in dusty regions.
}
\end{figure*}

At low Galactic latitudes, extinction increases rapidly and the color of the blue tip of the stellar locus becomes dramatically redder as fainter magnitudes are probed (Fig.~\ref{fig_blueplot-dist}).  In these dusty regions, reddened blue tip stars may be redder than intrinsically red foreground stars.  Moreover, blue tip stars in these regions are spread out along the reddening vector, making it hard to isolate stars of a given reddening and distance using $D$.  We attempt to identify and exclude such regions when analyzing blue tip maps.

\subsection{Blue Tip Maps}
\label{subsec:bluetipmaps}

\begin{figure*}[!tb]
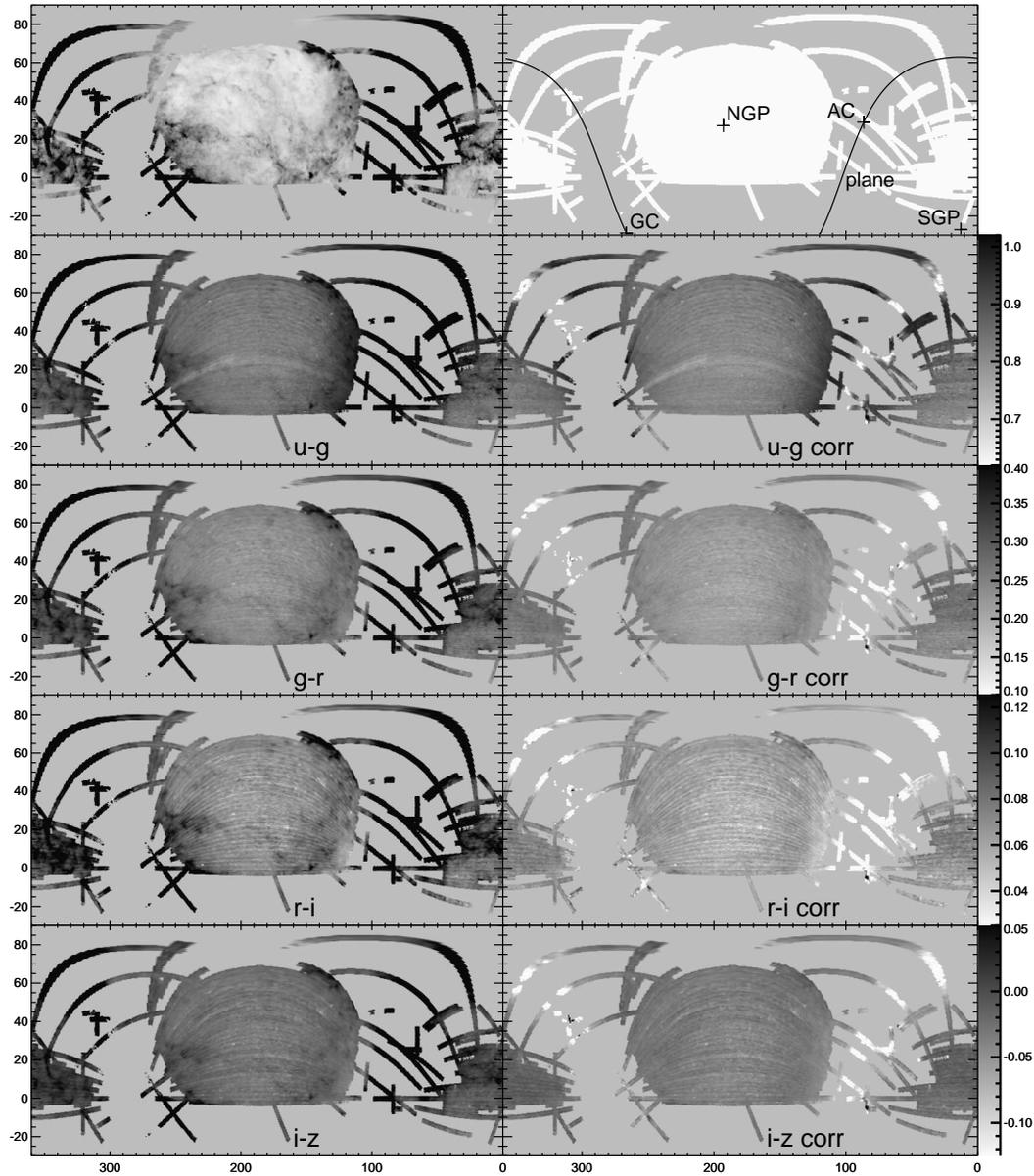

\dfplot{bluemaps.ps}
\figcaption{
\label{fig_bluemaps}
Blue tip colors over the SDSS footprint, for each SDSS color.  The left hand panels give the observed blue tip color, while the right hand panels give blue tip colors after correction for dust extinction according to SFD and the coefficients $(A_a-A_b)/(\mathrm{SFD}~E(B-V))$ presented in this paper.  The upper row of panels gives SFD and the locations of important points in the sky.
}
\end{figure*}

We measure the blue tip color over the entire SDSS footprint, in each color and for a variety of ranges of $D$ (Fig.~\ref{fig_bluemaps}).  Here we present maps using blue tip fits to stars in a broad range of magnitudes: $10 < D < 19 \MAG$, using fits to the stars in 10 adjacent SDSS fields.

\begin{figure*}[tbh]
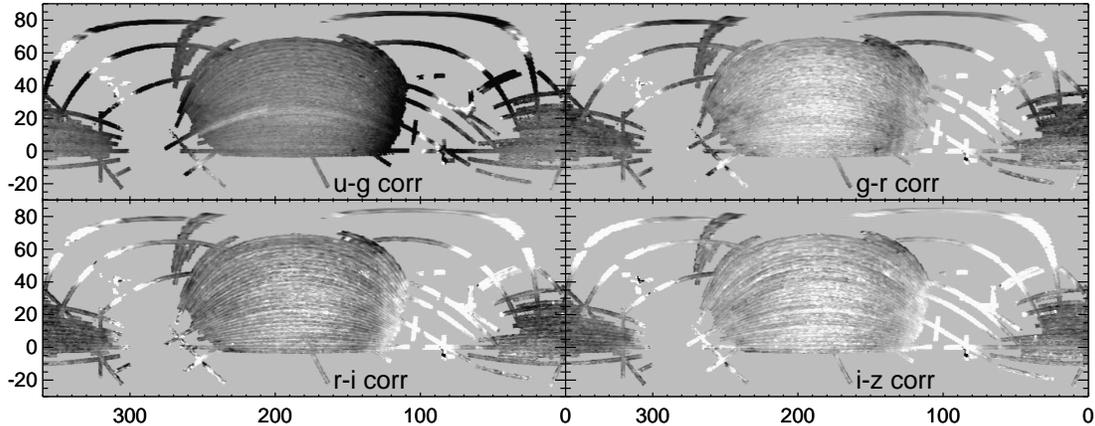

\dfplot{bluemaps-resid.ps}
\figcaption{
\label{fig_bluemaps-resid}
Stretched version of Fig.~\ref{fig_bluemaps}, highlighting residuals.  The Monoceros stream is prominent at the right hand side of the NGP region of the footprint.  White to black spans 0.225, 0.11, 0.05, and 0.06\MAG.
}
\end{figure*}

The maps show unmistakable signatures of dust (dark clouds in left panels of Fig.~\ref{fig_bluemaps}) that are almost entirely removed when stellar colors are corrected according to the SFD map.  The residuals at high Galactic latitude after extinction correction are dominated by problems with the survey calibration---striping in the SDSS scan direction and a few runs with bad zero points in $u-g$ (Fig.~\ref{fig_bluemaps-resid}).  There is also a slowly varying residual in which blue tip color becomes redder near the Galactic plane, until deep in the Galactic plane when the color shifts very blue.  

The former effect is caused by the various stellar population sampled in different parts of the sky.  In the plane, the higher metallicity, redder disk population becomes increasingly dominant over the halo population.  The latter effect is a result of the violation of the assumption that all of the stars are behind all of the dust.  Stars in front of the dust are dereddened with the full SFD dust column, and are therefore rendered extremely blue.  These stars are then seen by the blue tip algorithm as the signature of the blue tip, resulting in a spuriously blue color.  Even in less dusty regions where at least some MSTO stars are behind all of the dust, the method can fail if an insufficiently faint range of $D$ is used.  In this case, MSTO stars behind different amounts of dust are grouped together in finding the best fit, blurring the blue edge of the stellar locus.  Because $\sigma_P$ for the fit is fixed, this will result in a bluer than average blue tip color, while SFD would track the color of the MSTO stars behind all of the dust---the reddest stars in the group.  These effects, however, are only noticeable in the dustiest region of the sky, where $|b| \lesssim 15\degree$.

The Monoceros stream is clearly visible as a feature at $(\alpha,\delta) = (110\degree, -10\degree)$ -- $(110\degree, 60\degree)$ in the blue tip maps, because its MSTO is bluer than that of the thick and thin disks \citep{Newberg:2002}, except in $u-g$.

There are two types of survey-systematic induced residuals in the maps, both of which are seen as striping along the SDSS scan direction.  The less important are the few, large, $\sim 3\degree$ wide stripes that are most noticeable on the eastern edge of the $u-g$ blue tip maps.  These regions correspond to the footprints of SDSS runs that have slight zero point offsets from the rest of the survey.  The more troubling residual is the universal small-scale striping along the scan direction.  This corresponds to the 13 arcminute scale of the SDSS camera columns.  We have not fully examined this residual, but attribute it to different filter responses between the 6 SDSS camera columns \citep{Doi:2010}.  The SDSS ubercalibration algorithm ties stars of some mean color together, but does not account for color terms between the SDSS camera columns \citep{Padmanabhan:2008}.  As the MSTO stars are substantially more blue than the typical SDSS star, they are particularly vulnerable to color terms.  Complicating this explanation of the camera column striping is the observation that the best fit camera column offsets derived in \textsection \ref{subsec:globalfit} are different for the brightest range of $D$ used than for fainter ranges of $D$, which are mutually compatible.  We interpret this as a saturation or nonlinearity effect that depends on camera column, though we remove any stars flagged as saturated from the analysis.  These systematics need to be addressed in fits to maps of the blue tip.

\section{Fitting the Blue Tip Map}
\label{sec:fittingthemap}

The sensitivity of the blue tip colors to dust reddening makes for a natural test of the SFD map.  We carry out the simplest tests of SFD permitted by the blue tip maps here, checking the dust map normalization and the $R_V = 3.1$ reddening law assumed by SFD for extrapolating reddening to colors other than $B-V$.  The reddening law used in SFD is from \citet{O'Donnell:1994}, which is similar to the \citet[CCM]{Cardelli:1989} reddening law.

The SFD dust map is often used as a map of reddening $E(B-V)$, but is based fundamentally on thermal emission from dust.  In SFD, the thermal emission is converted to reddening based on a single dust map normalization constant.  This constant was derived by comparing the observed reddenings $E(B-V)$ of a sample of elliptical galaxies to the SFD temperature-corrected emission from dust at their locations \citep{Faber:1989}.

We repeat this test, using the colors of the blue tip of the stellar locus in place of the $B-V$ color of galaxies.  Because of the five band photometry provided by the SDSS, we are able to extend the SFD test by additionally checking the assumed O'Donnell $R_V=3.1$ reddening law.  Because of the number and accuracy of the blue tip color measurements, we are able to further look for spatial variation in the SFD normalization constant and the reddening law.  The method is ultimately the same as in SFD, however: we find the best fit normalization constants that convert from dust emission (or $E(B-V)_\mathrm{SFD}$, as these are proportional) to reddening.  We extend this test to multiple colors and check for variation over the sky.

To perform this test each blue tip map would be fit, ideally, as 
\begin{equation}
(m_a-m_b)_\mathrm{bluetip} = R_{a-b} E(B-V)_\mathrm{SFD} + C, 
\end{equation}
over a part of the sky, where $(m_a-m_b)_\mathrm{bluetip}$ is the blue tip map for the color $a-b$, $R_{a-b}$ is the normalization constant to be measured, and $C$ is the intrinsic blue tip color.  However, because the blue tip of the stellar locus is not a universal color standard, but rather varies with position in a way likely substantially covariant with the dust map, this fit is impractical.  Moreover, the survey striping artifacts discussed in \textsection \ref{subsec:bluetipmaps} could further throw off the fit.  To account for these two effects, we instead fit 
\begin{equation}
(m_a-m_b)_\mathrm{bluetip} = R_{a-b} E(B-V)_\mathrm{SFD} + Q_r(f) + C_i
\end{equation}
where $Q_r(f)$ is a quadratic for run $r$ in SDSS field number $f$ and $C_i$ is the camera column offset for camera column $i$.  We fix $C_1$ to be zero to remove the degeneracy with the constant term in $Q_r(f)$.  The quadratic $Q_r(f)$ simultaneously accounts for zero point errors in the SDSS calibration and slow intrinsic variation in blue tip color.

Because we bin 10 SDSS fields together for each blue tip measurement, variations in the dust map on scales much smaller than 1.5$\degree$ in the scan direction will not be captured in the blue tip measurements.  The blue tip method is not a linear operator on the underlying stellar colors, so the measured color of the blue tip is not linearly related to the mean of $E(B-V)_\mathrm{SFD}$ in the fields.  In this work, for each set of fields contributing to a blue tip measurement, we use the median $E(B-V)_\mathrm{SFD}$ in those fields as proportional to the expected reddening of that measurement.  This is not strictly correct, but insofar as the dust does not vary too quickly, this filtering seems to approximate the correct behavior.

After finding the best fit $R_{a-b}$ in each color, we repeat the blue tip measurements on the SDSS stars, this time with individual stellar colors corrected according to the derived fit parameters $R_{a-b}$, at the full resolution of the dust map.  This process converges in a few iterations.  The iteration renders the final fits insensitive to the details of the dust map filtering.  We have verified the insensitivity to the filtering by noting that fits to the unbinned blue tip maps using an unfiltered SFD map give fit coefficients that are compatible with the fits to the binned blue tip maps.

\begin{figure}[tbh]
\dfplots{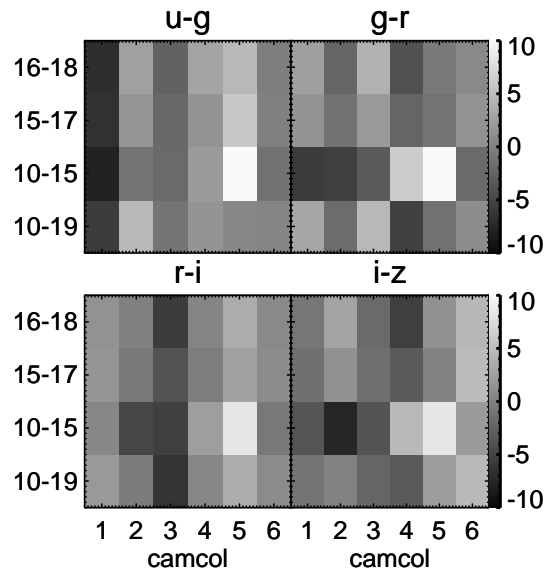}
\begin{center}
\end{center}
\figcaption{
\label{fig_blueadjustplot}
The camera column offsets used to destripe the SDSS blue tip maps, derived from the global blue tip map fits (\textsection \ref{subsec:globalfit}).  The left-hand axis labels give the range of $D$ to which the offsets apply.  The right hand labels give the value of the grayscale in$\MMAG$.  The offsets are compatible for different ranges of $D$, except at the brightest magnitudes ($10 < D < 15$), where nonlinearity effects may be coming into play.
}
\end{figure}

We model the striping in camera columns by constant terms in the fit for each camera column.  Empirically this correction is satisfactory, but we note that if our interpretation of the camera column offsets as derived from color terms is correct, the response of each camera column to dust is slightly biased.  However, we detect camera column offsets of $\sim 5\MMAG$ over differences in stellar colors of a few tenths of magnitudes, implying that this bias is less than 1\%  (Fig.~\ref{fig_blueadjustplot}).  Furthemore, the bias will be different in each camera column and it must be, on average, zero.  We have further confirmed that our results are unaffected by the color terms by separately fitting the dust map to each SDSS camera column individually, obtaining fits consistent with the global 6 camera column fit (\textsection \ref{subsec:regions}).

\subsection{Fits to Individual SDSS Runs}
\label{subsec:constrainedruns}

\begin{figure}[tbh]
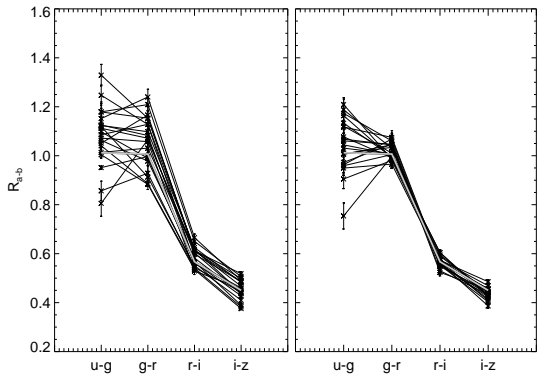

\dfplot{rv_runs_spectrum.ps}
\figcaption{
\label{fig_rv_runs_spectrum}
$R_{a-b}$ for the four SDSS colors, for a selection of individually well constrained SDSS runs.  The reddening spectrum has a similar shape among the different runs, but there is substantial scatter in the overall normalization (left).  After forcing the mean of $R_{g-r}$, $R_{r-i}$, and $R_{i-z}$ to match for each run, there is close agreement among the various dust extinction spectra (right).  Error bars account only for the formal statistical uncertainties.  The thick gray line in each plot gives the global best fit $R_{a-b}$ derived in this work.
}
\end{figure}

\begin{figure*}[tbh]
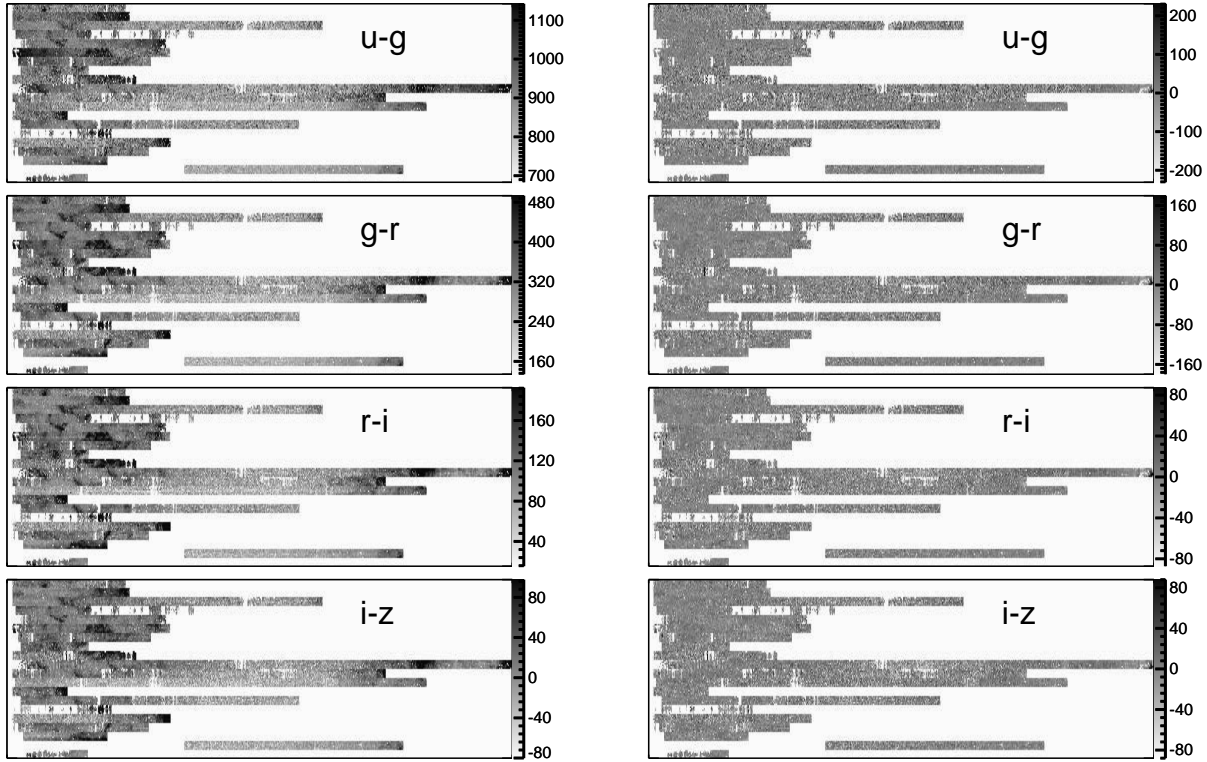

\begin{center}
\dfplottwo{rv_runs_raw.ps}{rv_runs_resid.ps}
\end{center}
\figcaption{
\label{fig_rv_runs}
Blue tip colors (left) and residuals (right) for the runs in Fig.~\ref{fig_rv_runs_spectrum}, in$\MMAG$.  Coherent signals from the dust are manifest (left).  After fitting according to \textsection \ref{sec:fittingthemap}, the residuals show little structure (right).
}
\end{figure*}

In some cases, a single run is long enough and dusty enough to individually constrain the dust reddening coefficients $R_{a-b}$ in each color at the 10\% level, even when fitting for camera column offsets and a quadratic.  The fits show substantial variation in $R_{a-b}$ (Fig.~\ref{fig_rv_runs_spectrum}).  Most of this variation takes the form of an overall difference in dust absorption normalization, rather than as variation in the shape of the dust reddening spectrum.  Some low, outlying values of $R_{u-g}$ suggest that the run is so extinguished that substantial numbers of stars are not detected in the $u$ band, ruining the fit.  The fits remove nearly any trace of dust in these runs (Fig.~\ref{fig_rv_runs}).

Fits to the runs individually well constrained by the dust map indicate significant variation in normalization ($\sim 20\%$) as well as, to a lesser extent, variation in dust extinction spectrum.  Accordingly, in finding a global fit to the dust properties, we remove from the fit regions that we deem discrepant from the ``average'' dust properties.  To do this, first we individually fit each SDSS run.  Runs with best fit $R_{a-b}$ more than $5\sigma$ from the inverse variance weighted average of $R_{a-b}$ for all the SDSS runs are excluded from the global fit.  Additionally, the global fit is iterated and $\sigma$-clipped field-by-field at $5\sigma$.

\subsection{Global Fit to All SDSS Runs}
\label{subsec:globalfit}

We find the best fit (least squares sense) combination of the SFD dust map, constant offsets for each camera column other than the first, and quadratics in field number for each run to the blue tip map via singular value decomposition.  We do not fit for the offset for the first camera column because changing the offsets for all of the camera columns is degenerate with changing the constant offsets in the quadratics for each run.  Specifically, we find the parameters $\mathbf{x}$ satisfying 
\begin{equation}
\mathbf{A}^{\intercal}\mathbf{C}^{-1}\mathbf{A}\mathbf{x} = \mathbf{A}^{\intercal}\mathbf{C}^{-1}\mathbf{b}
\end{equation}
which provides the least squares solution to $\mathbf{A}\mathbf{x} = \mathbf{b}$.  Here $\mathbf{C}$ is a diagonal covariance matrix, with the variances determined by the formal statistical uncertainties in the blue tip fit to each field (or binned fields).  The blue tip color for each field (or binned fields) is in $\mathbf{b}$.  The design matrix $\mathbf{A}$ has dimensions $n_{\mathrm{field}} \times n_{\mathrm{param}}$.  It contains one column for the SFD dust map values in each field, five columns for the offsets for the SDSS camera columns two through six, and three columns for each of the 792 runs composing the SDSS-III, for the three terms in the quadratics for each run.  

The SDSS-III contains imaging data on 1,147,506 fields.  We require that the SDSS score\footnote{The SDSS score of a field reflects sensitivity to point sources and is derived from sky brightness and seeing.  It further is capped at 0.5 for fields deemed unphotometric or runs fields using binned pixels (i.e., Apache Wheel fields).} of each run be greater than 0.5 and the PSF FWHM be less than 1.8 arcseconds in $r$, which reduces the number of fields to 686,554.  Excluded runs typically are unphotometric, have bad seeing or are Apache Wheel calibration runs.  The blue tip method is run on all of the remaining fields, or binned sets of these fields.  It occasionally fails, due to an insufficient number of stars or apparent detection of the blue tip blueward of $-0.75\MAG$ or redward of 2.4, 2.4, 1.8, and 1.6$\MAG$ in $u-g$, $g-r$, $r-i$, and $i-z$, respectively.  The number of binned fields that successfully make it through the blue tip fit depends on the range of $D$ used and the color, but in the ranges of $D$ examined here it ranges from 684,000 to 677,000.  Unbinned fields fail for lack of stars at high Galactic latitudes, and so fewer fields are successful: in ranges of $D$ of interest, between 683,000 and 620,000 fields pass.  The median number of stars contributing to the fits in each unbinned field at $10 < D < 19$ is 100.  About $10^8$ stars enter into the fits.

\begin{figure*}[tbh]
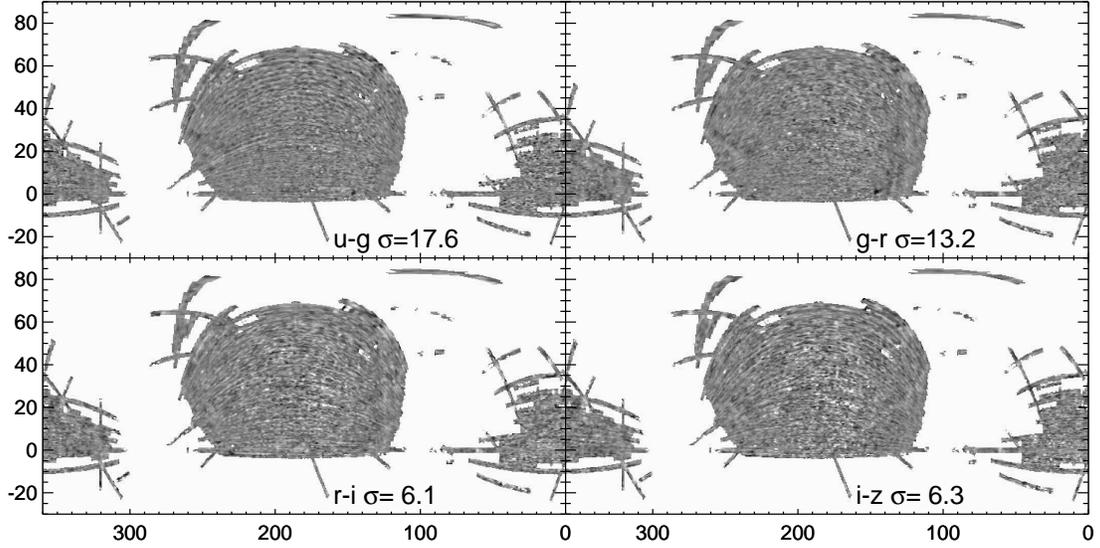

\dfplot{bluemaps-fitresid.ps}
\figcaption{
\label{fig_bluemaps-fitresid}
Blue tip color residuals after removing best fit linear combination of dust map and polynomials, as described in \textsection \ref{subsec:globalfit}.  White corresponds to $+3\sigma$, while black corresponds to $-3\sigma$.
}
\end{figure*}

The global fit produces the coefficients $R_{a-b}$ in Table~\ref{table:rab} and at high latitudes leaves few noticeable coherent residuals (Fig.~\ref{fig_bluemaps-fitresid}).  Nevertheless, a few dust clouds clearly are imperfectly subtracted.  Most prominently, in the north, a few undersubtracted clouds in $g-r$ stand out on the eastern side of the north Galactic cap (NGC).  Undersubtracted clouds in the northwest and at ($\alpha$, $\delta$) = (145\degree, 0\degree) cause the runs including them to be excluded from the fit.  The former feature is the largest residual in the \citet{Peek:2010} maps.  Subtraction in the south is remarkably clean, though at right ascensions and declinations of (340\degree, 20\degree) and (60\degree, 0\degree) there are slightly oversubtracted clouds.  A cloud at (45\degree, 20\degree) barely makes it into the SDSS footprint, and is badly oversubtracted, causing the exclusion of its run from the global fit.  In $g-r$ especially, a residual associated with the Monoceros stream stands out at the western edge of the NGC.  We have tried masking this region and repeating the fit, but as the Monoceros stream is not correlated with the dust, the fit results were unchanged.

\begin{deluxetable}{cccc}
\tablewidth{0pc}
\tablecaption{$R_{a-b}$ for the SDSS colors}
\tablehead{
\colhead{SDSS color} & \colhead{SFD} & \colhead{$R_V = 3.1$ O'Donnell} & \colhead{blue tip}
}
\startdata
$u-g$ & $1.362$ & $1.138$ & $ 1.01 \pm  0.10$ \\
$g-r$ & $1.042$ & $1.141$ & $ 1.01 \pm  0.08$ \\
$r-i$ & $0.665$ & $0.616$ & $ 0.57 \pm  0.05$ \\
$i-z$ & $0.607$ & $0.624$ & $ 0.45 \pm  0.05$
\enddata
\tablecomments{
\label{table:rab}
$R_{a-b}$ according to the original SFD prescription, an $R_V = 3.1$ O'Donnell prescription with updated SDSS filter definitions, and the best blue tip global fit coefficients.  The formal statistical error bars are negligible; those presented here are based on the empirical field-to-field variation in best fit extinction law as discussed in \textsection \ref{sec:discussion}.  This gives highly covariant uncertainties because the field-to-field variations are dominated by changes in best fit dust map normalization (see the covariance matrix in Table~\ref{table:cov}).
}
\end{deluxetable}

\begin{figure}[!tbh]
\dfplots{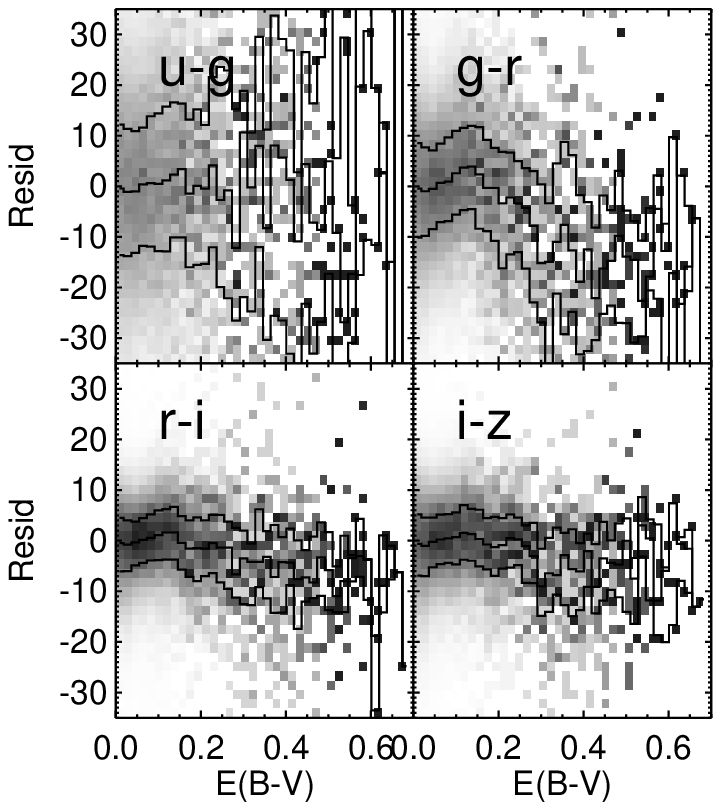}
\figcaption{
\label{fig_extresid}
Blue tip color residuals with $\EBVSFD$.  The residuals in $u-g$, $r-i$, and $i-z$ reassuringly show no trend with $E(B-V)$.  In $g-r$, the residuals have some structure, but the positive slope at $E(B-V) < 0.2$ and the negative one around $0.2 < E(B-V) < 0.4$ are only 5\% effects---the residuals are largely uncorrelated with $E(B-V)$.
}
\end{figure}

The blue tip color residuals as a function of $\EBVSFD$ are generally flat (Fig.~\ref{fig_extresid}).  In $r-i$ and $i-z$, any trend with $E(B-V)$ is at less than the 2\% level.  In $u-g$ there is no discernible trend, though the scatter is large.  In $g-r$ there seem to be disturbing 5-10\% trends with $E(B-V)$ but they may stabilize around $E(B-V) = 0.4\MAG$, with reddenings off by only 15$\MMAG$ there.

\begin{figure}[!tbh]
\dfplots{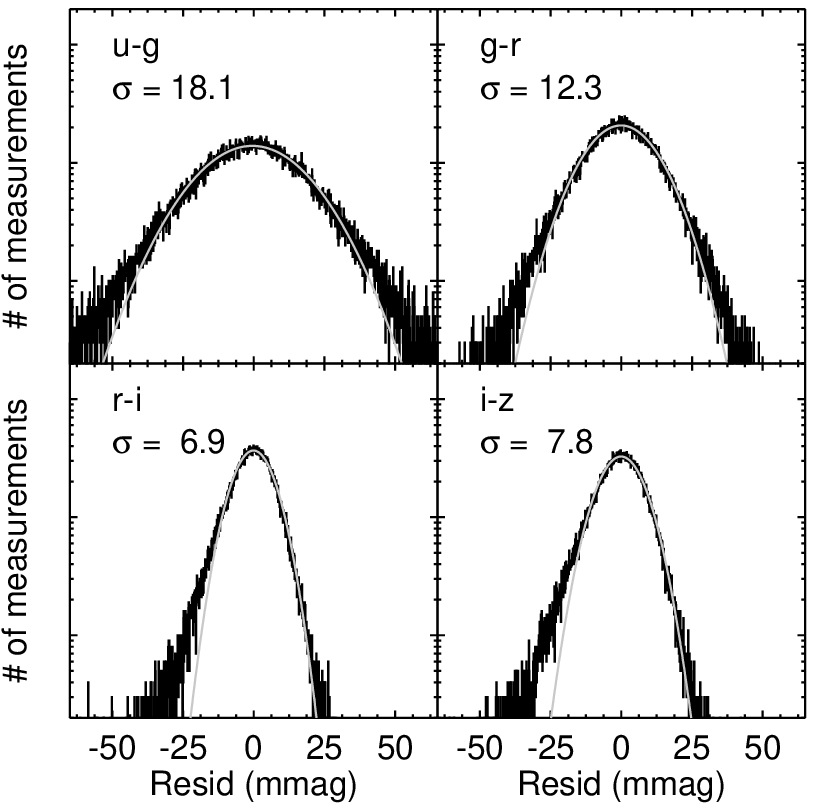}
\figcaption{
\label{fig_resid}
Fit residuals, in$\MMAG$.  The core of the distribution is well fit by a Gaussian, but the wings fall off more slowly than in a Gaussian.  The colors $r-i$ and $i-z$ have the tightest distributions, likely because the blue tip is less dependent on metallicity in these colors.
}
\end{figure}

The blue tip maps have residuals with distributions given in Fig.~\ref{fig_resid}.  The core of the distribution is Gaussian, but the wings are non-Gaussian, falling off more slowly than would be expected.   The blue tip method gives median estimated uncertainties as 17.5, 12.5, 5.7, and 5.8$\MMAG$, closely reproducing the Gaussian fits to the residual distribution, which give 18.1, 12.3, 6.9, and 7.8$\MMAG$ in $u-g$, $g-r$, $r-i$, and $i-z$, respectively.

\begin{figure*}[tbh]
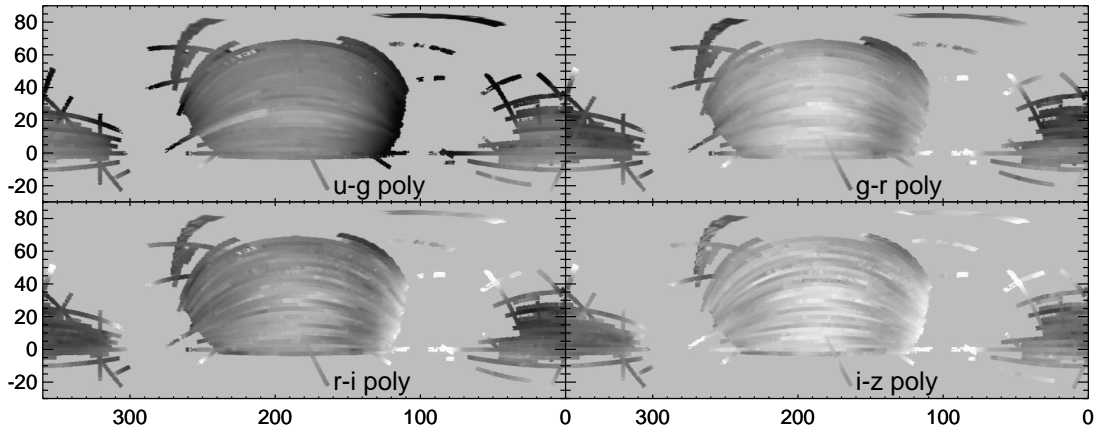

\dfplot{polyfits.ps}
\figcaption{
\label{fig_polyfits}
Map of the polynomial terms from the global fits.  The scale is the same as in Fig.~\ref{fig_bluemaps-resid}.  These maps model the intrinsic variation of the blue tip color on the sky as well as per-run SDSS calibration offsets.
}
\end{figure*}

The polynomial terms from the global fit are intended to model the slow, intrinsic variation in the blue tip color and per-run calibration offsets, and so are of independent interest (Fig.~\ref{fig_polyfits}).  These maps are clearly heavily striped.  Some of this striping is clearly removing calibration problems with the SDSS, as in the eastern side of the NGC in $u-g$.  The striping at high latitudes in the north in other colors is probably substantially an artifact of the low signal in these regions, but the typical stripe-to-stripe difference there is only $\sim 5\MMAG$, and at worst $10 \MMAG$.  

The SDSS imaging contains a number of runs crossing the Galactic plane, as part of SEGUE \citep{Yanny:2009}.  All of these runs and a few other low Galactic latitude runs get identified as discrepant by the fitting algorithm and discarded from the final fit.  In a run-by-run treatment, this is inevitable, as at very high extinction the blue tip method will fail because the stars are not behind all the dust.   However, for $|b| > 10\degree$, inspection of the blue tip residual maps suggests that the SEGUE runs could be well fit also, although we have not tried that here.  The signal from the dust in the SEGUE runs is sufficiently rich to merit treatment separate from the bulk of the high Galactic latitude sky, as is done in Berry \etal (in prep).

\subsection{Fits to different sky regions}
\label{subsec:regions}

\begin{figure*}[tb]
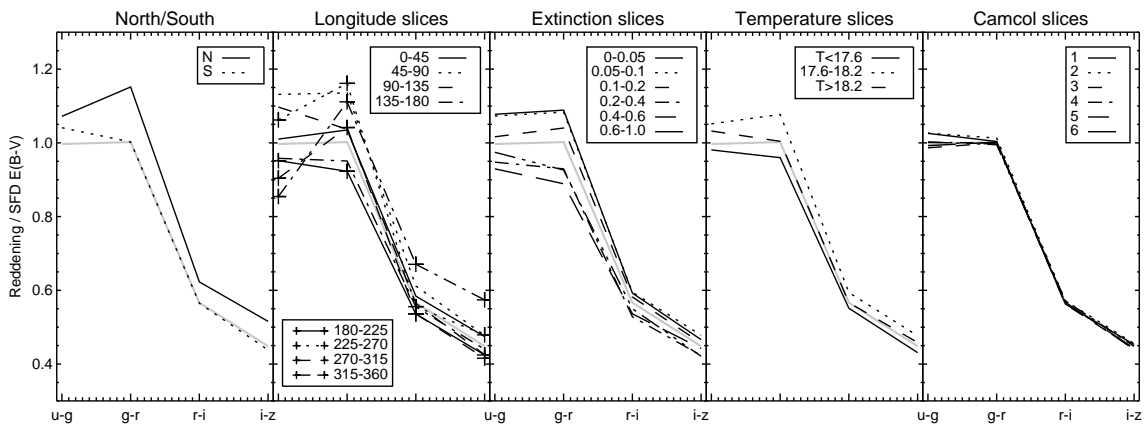

\dfplot{dustskysubset.ps}
\figcaption{
\label{fig_dustskysubset}
$R_{a-b}$ for the SDSS colors over different subsets of the SDSS footprint.  The first panel gives the coefficients for the north ($b>25\degree$) and south ($b<-25\degree$).  The next panel gives the best fit coefficients for eight slices in Galactic longitude.  The third panel gives coefficients in regions of sky with increasing extinction.  The fourth panel divides the sky into regions of three different temperatures, and the fifth panel divides the survey by SDSS camera column.  The thick gray line in each panel gives the global best fit $R_{a-b}$ derived in this work.
}
\end{figure*}

Given the presence of dust-correlated residuals in different parts of the sky, one wonders whether the fit is actually ``global'' in a useful sense, or is primarily a product of the footprint we have decided to look at.  Accordingly, we have cut the sky into a number of subregions and separately found the best fit $R_{a-b}$ in each.  To test for large scale variations in $R_{a-b}$, we divide the sky into northern ($b>25\degree$) and southern ($b<-25\degree$) Galactic regions, as well as octants of the sky divided by lines of constant Galactic longitude (Fig.~\ref{fig_dustskysubset}).  To try to test variation in dust properties as a function of extinction or temperature, we divide the sky into regions of dust of different extinctions and temperatures.  Finally, to verify that we have satisfactorily accounted for the camera column offsets, we divide the SDSS survey into its six camera columns.

We restrict to the main global fit region and apply the globally derived camera column offsets to the blue tip colors.  We remove best fit polynomials from the filtered SFD dust map and from the blue tip colors, as in \textsection \ref{sec:fittingthemap}, using the global fit region.  We then find the best fit of the filtered, polynomial-subtracted dust map to the polynomial-subtracted blue tip map, restricted to various subsets of the full region.

The results are encouragingly consistent, given that we have already seen that dust reddening normalization can vary substantially, along with, to a lesser extent, its spectrum (Fig.~\ref{fig_rv_runs_spectrum}).  We find the dust extinction normalization and spectrum to be consistent at the 10\% level.  The north/south normalization discrepancy is the most surprising, with the north preferring a normalization about 15\% larger.  We also find that dust with $0.6 < E(B-V) < 1.0\MAG$ prefers a $\sim 15\%$ smaller normalization than dust with $0 < E(B-V) < 0.05\MAG$.  Dust at different temperatures likewise has similar extinction normalizations and spectra.

\subsection{Fits to a Mock Star Catalog}
\label{subsec:mockcatalog}

The reliability of the fitting procedure can be tested by running the fit on a star catalog from a mock galaxy model.  The mock catalog was generated with the {\em galfast} catalog generation code (Juri\'{c}, in prep).  Given a model of stellar number density, metallicity, a 3D extinction map, the photometric system, and instrumental errors, {\em galfast} generates realistic mock photometric survey catalogs.

As inputs, we used the number density distribution parameters from \citet{Juric:2008} and the metallicity distribution from \citet{Ivezic:2008}. The three-dimensional dust distribution map was generated using the model of \citet{Amores:2005}. The generated $u$, $g$, $r$, $i$, and $z$ magnitudes were convolved with magnitude-dependent errors representative of SDSS, and the final catalog was flux limited at $r<22.5$.

The three-dimensional dust map was modified from \citet{Amores:2005} to contain the small-scale clouds seen in SFD.  This was performed by scaling each line of sight from the model by the ratio $E(B-V)_\mathrm{SFD}/E(B-V)_\mathrm{model, 100 kpc}$, so that the model matches SFD at 100 kpc.  The resulting map contains the expected average three dimensional distribution of the dust combined with the angular structures present in SFD.

We construct a mock galaxy using \emph{galfast} and then construct catalogs of observations of these stars for each SDSS run.  The resulting catalogs are proccesed by the blue tip analysis code in exactly the way that the actual SDSS catalogs are processed.  As for the actual catalogs, the median number of stars per field is $\sim 100$.

Blue tip fits performed on the mock galaxy recover the $R_{a-b}$ used to within 3\%.  This verifies that we properly account for shifts in the blue tip color due to metallicity.  However, for these tests we have assumed that SFD correctly predicts the line of sight column density and that the reddening law is the same everywhere on the sky, at least at high Galactic latitudes and in the optical.  Insofar as the final fit residuals are largely flat (Fig.~\ref{fig_bluemaps-fitresid}), this assumption seems justified.  Moreover we can verify that the ratios of the $R_{a-b}$ derived independently from SFD agree with the ratios of the $R_{a-b}$ derived here.

\subsection{Fitting Ratios of $R_{a-b}$ without SFD}
\label{subsec:btbt}

The preceding analysis has assumed that SFD is a good template for the dust.  However, we can relax our dependence on SFD if we restrict our attention to ratios between different $R_{a-b}$.

\begin{figure*}[!tbh]
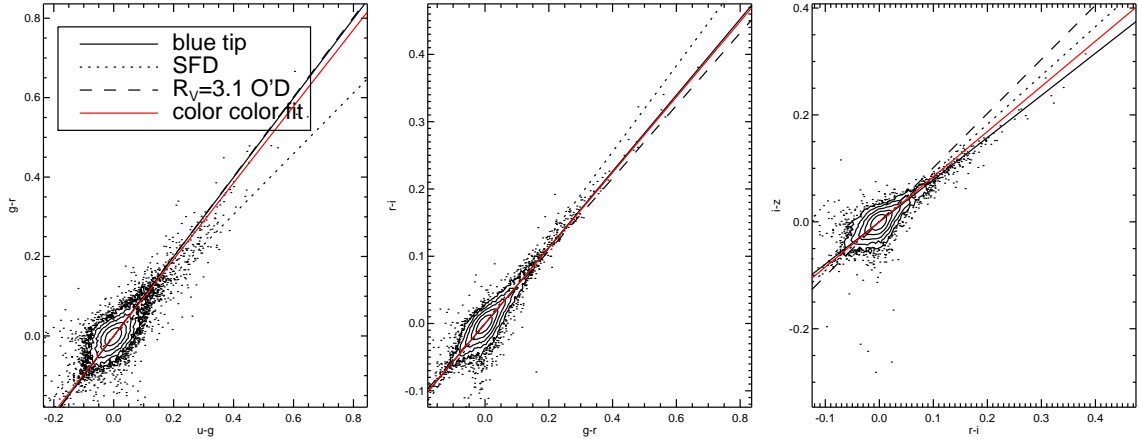

\dfplot{btbtfit.ps}
\figcaption{
\label{fig_btbt}
Plots of the colors of the blue tip in different colors, for camera column one.  Spatially slowly varying terms have been removed, so points should fall along the line $R_{a-b}/R_{c-d}$.  The reddening coefficients presented here best trace the distribution of points.
}
\end{figure*}

In the absence of changing stellar populations and calibration errors, blue tip colors will fall along a single line given by the reddening vector.  By removing a quadratic in field number and accounting for camera column offsets as in \textsection \ref{subsec:globalfit}, we remove variations in blue tip color not associated with reddening.  By fitting a line to the resulting polynomial-subtracted blue tip color-color diagrams, we measure the ratio of the $R_{a-b}$.  Some striping along the SDSS scan direction is evident in the blue tip residuals despite our attempts to remove it, and this striping will be deleterious to the fit results because we cannot rely on its being uncorrelated with the dust template we fit.  Accordingly, we fit lines for each camera column, eliminating the effect of striping, and average the results (Fig.~\ref{fig_btbt}).

These color-color fit results are in good agreement with the results of \textsection \ref{subsec:globalfit} (Table~\ref{table:btbt}).  The results agree to about 1$\sigma$ with the SFD fit results, where the uncertainties $\sigma$ are given by the standard deviations of the fit results between the camera columns.  The original SFD prescription is seen to clearly overestimate $R_{u-g}/R_{g-r}$ (see \textsection{\ref{subsubsec:rvnglobal}}), while both the original SFD prescription and the updated $R_V=3.1$ O'Donnell reddening law are found to underestimate $R_{r-i}/R_{i-z}$.

\begin{deluxetable}{ccccc}
\tablewidth{0pc}
\tablecaption{Blue Tip Color-Color Fit Results}
\tablehead{
\colhead{Color ratio} & \colhead{SFD} & \colhead{O'Donnell} & \colhead{blue tip} & \colhead{color color}
}
\startdata
$A_{u-g}/A_{g-r}$ & 1.307 & 0.997 & 1.005 & $1.019 \pm 0.025$ \\ 
$A_{g-r}/A_{r-i}$ & 1.567 & 1.852 & 1.776 & $1.728 \pm 0.038$ \\ 
$A_{r-i}/A_{i-z}$ & 1.096 & 0.987 & 1.260 & $1.237 \pm 0.042$ \\ 
\enddata
\tablecomments{
\label{table:btbt}
Blue tip color-color fit results.  The results are in close agreement with the global blue tip fits, while excluding the original SFD reddening law and an $R_V=3.1$ O'Donnell reddening law.  These fits do not use SFD as a template and so test the shape of the reddening law without reference to SFD.
}
\end{deluxetable}

\section{Discussion}
\label{sec:discussion}

In this section, we use the blue tip fits presented above to constrain reddening laws and to test the SFD normalization and temperature correction.  We then briefly point out a discrepancy in the color of the dereddened blue tip between the north and south, which is plausibly a consequence of a different stellar population in the north than in the south.

\subsection{Constraining Reddening Laws}
\label{subsec:fitrv}

\subsubsection{Connecting Reddening Laws to $R_{a-b}$}
\label{subsubsec:rvn}

A reddening law gives the extinction $A(\lambda)$ over a range of wavelengths $\lambda$ and so makes a prediction for the $R_{a-b}$ that we have fit.  The extinction in band $b$, in the limit that the variation in the extinction over that band in small, is given by $\Delta m_b = \Aeff{b} E(B-V)$, with
\begin{equation}
\lambda_{\mathrm{eff},b,S} = \frac{\int{d\lambda \lambda S(\lambda) W_b(\lambda)}}{\int{d\lambda S(\lambda) W_b(\lambda)}}
\end{equation}
Here the system response for band $b$ is given by $W_b(\lambda)$, $S(\lambda)$ is the source spectrum in photons/s/\AA, and $A(\lambda)$ is the reddening law, normalized to give $\Aeff{B}-\Aeff{V} = 1$.  The reddening in the color $a-b$ is then given by $(\Aeff{a}-\Aeff{b}) E(B-V)$, from which it follows that $R_{a-b} = (\Aeff{a}-\Aeff{b}) E(B-V)/\EBVSFD$.  Here we have assumed that the variation in the extinction over the band pass is small; for $E(B-V) < 1$, this assumption changes the predicted $R_{a-b}$ by less than 1\% in $r-i$ and $i-z$, and less than 5\% in $g-r$ and $u-g$.

Accordingly, we evaluate the extinctions in each SDSS pass band using the SDSS system throughputs and a MSTO source spectrum from a Kurucz model with $T_{\mathrm{eff}} = 7000\mathrm{K}$, for a variety of reddening laws \citep{Gunn:1998, Kurucz:1993}.  We examine in particular the CCM, O'Donnell, and \citet[F99]{Fitzpatrick:1999} reddening laws, which are parameterized by $R_V = A_V/E(B-V)$.  We find the best fit factors $R_V$ to our measured $R_{a-b}$ for each of these reddening laws.

The other free parameter in these fits is the best fit normalization $N^\prime = E(B-V)/\EBVSFD$.  We do not report this parameter directly.  The SFD dust map is based on a map of thermal emission from dust and is correspondingly proportional to $\tauum{100}$, the optical depth of dust at 100$\mum$.  Empirically, the ratio between $\tau_{100\mum}$ and $E(B-V)$ is itself a function of $R_V$, and so $N^\prime$ is covariant with $R_V$.  However, the ratio between $\tauum{1}$ and $\tauum{100}$ depends less on $R_V$, so we report $N = A_{1\mum,~\mathrm{predicted}}/A_{1\mum,~\mathrm{SFD}}$.  Here $A_{1\mum,~\mathrm{predicted}}$ is the extinction at $1\mum$ implied by the best fit reddening law and $A_{1\mum,~\mathrm{SFD}}$ is the SFD-predicted extinction, extrapolated to $1\mum$ following the $R_V=3.1$ O'Donnell reddening law assumed by SFD.  The quantity $A_{1\mum,~\mathrm{SFD}}$ is simply $1.32 \cdot \EBVSFD$.  We find that the F99 reddening law provides the best fit to the $R_{a-b}$ we measure, so we additionally mention for reference that the F99 reddening law ratio $E(B-V)/A_{1\mum}$ is 11\% larger than the O'Donnell ratio $E(B-V)/A_{1\mum}$, though this depends on the adopted $R_V$ and source spectrum.

\subsubsection{$R_V$ and $N$ for Individual Runs}
\label{subsubsec:rvnindrun}

\begin{figure}[tbh]
\dfplots{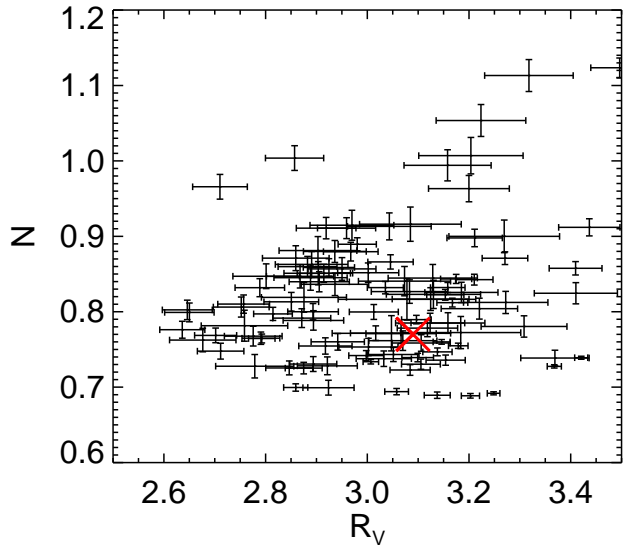}
\figcaption{
\label{fig_rv_runs_rvnorm}
Best fit $R_V$ and dust map normalization $N$ for SDSS runs that individually well constrain $R_{a-b}$, examples of which are in Fig.~\ref{fig_rv_runs_spectrum}.  Dust map normalizations vary by as much as a factor of two, though most are consistent at the 20\% level.  The large red ``X'' indicates the location of the best fit $R_V$ and normalization to the global fit $R_{a-b}$.  Error bars account only for the formal statistical uncertainties.
}
\end{figure}

We have found that the dust extinction spectrum normalization and shape vary over the sky (Fig.~\ref{fig_rv_runs_spectrum}).  We can quantify this effect in terms of variation in $N$ and $R_V$ by fitting reddening laws to the runs that individually well constrain $R_{a-b}$ (\textsection \ref{subsec:constrainedruns}).  We use the F99 reddening law here rather than the traditional CCM or O'Donnell reddening law because the latter two reddening laws (especially the O'Donnell reddening law) predict that $R_{r-i} \approx R_{i-z}$, which is a poor fit to the data.  The derived $N$ and $R_V$ vary from run to run, especially as a $\sim 20\%$ scatter in normalization (Fig.~\ref{fig_rv_runs_rvnorm}).  Almost all of the runs are consistent with $2.8 < R_V < 3.2$.  The distribution of $R_V$ and $N$ validates our use of $1 \mum$ as a refererence wavelength for $N$, as the two parameters appear uncorrelated. 

\subsubsection{$R_V$ and $N$ for the Global Fit}
\label{subsubsec:rvnglobal}

In order to compute best-fit reddening laws for the global fit $R_{a-b}$, we need to account for the covariance in the $R_{a-b}$ over the sky in addition to the formal statistical uncertainties in the fit.  These uncertainties are about a part in one thousand.  However, the best fit normalization can vary at the 10\% level from field to field, and the part in a thousand uncertainty reflects averaging these fluctuations over many fields.  While the dust extinction spectrum seems relatively constant, it would be surprising if the average reddening reflected the reddening of any particular dust cloud at the part in a thousand level, so another method for estimating the uncertainties is needed.  The uncertainties reported in Table~\ref{table:rab} are the standard deviations in $R_{a-b}$ for 8 octants in Galactic longitude, and do not account for the covariance of the measurements in the different colors owing to the changing best fit normalization.

The consequence of adopting uncertainties based on the field-to-field variation in the dust extinction spectrum is that the resulting uncertainties on $N$ and $R_V$ will describe the range of $N$ and $R_V$ among the clouds within the footprint we analyze.  They do not give the uncertainties on the best fit $N$ and $R_V$ within our sample, which are more tightly constrained.  As mentioned, the formal statistical uncertainties are about one part in a thousand.  Uncertainties estimated from using different ranges of $D$ for selecting the stars give uncertainties of 2\%, 2\%, 1\%, and 2\% in $R_{a-b}$ for the colors $u-g$, $g-r$, $r-i$, $i-z$, respectively.  

The field-to-field variation in $R_{a-b}$ can be estimated from the sample covariance matrix of best fit $R_{a-b}$ in different sky regions.  We find the best fit $R_{a-b}$ at each point in the footprint.  Because over most of the footprint the signal-to-noise in the filtered blue tip maps is too low to make a reliable determination of $R_{a-b}$, we include nearby pixels in the fit with weights given by a 7$\degree$ smoothed Gaussian.  The best fit $R_{a-b}$ is then given by
\begin{equation}
R_{a-b, n} = \frac{\sum_i D_i C^{-1}_i W_{i, n} b_i}{\sum_i D_i^2 C^{-1}_i W_{i, n}}
\end{equation}
where $n$ is the pixel on the sky, $i$ indexes over blue tip measurements, $D$ gives the filtered dust map, $b$ and $C^{-1}$ give the blue tip measurement and its inverse variance, and $W_{i,n}$ is the weight matrix, corresponding to Gaussian smoothing with a FWHM of $7\degree$.  

The sample covariance matrix of the resulting maps of $R_{a-b}$ is used as the covariance matrix for fits of reddening laws to $R_{a-b}$ (Table~\ref{table:cov}).  Only pixels with estimated uncertainty in $R_{i-z}$ less than $0.01$ are used for computing the sample covariance matrix, to avoid overestimating the intrinsic variance due to the uncertainty in the estimates.  The largest eigenvalue of the covariance matrix ($2.4 \cdot 10^{-2}$) is 94 times larger than the smallest eigenvalue ($2.6 \cdot 10^{-4}$).  The eigenvector with the largest eigenvalue is roughly parallel to a vector corresponding to changing the normalization of the spectrum, and that with the smallest eigenvalue corresponds to changing the relative amount of reddening in $r-i$ to $i-z$.  This quantitatively agrees with our claim that the normalization of the reddening law is poorly constrained relative to its shape.

\begin{deluxetable}{ccccc}
\tablewidth{0pc}
\tablecaption{Eigenvalues and eigenvectors of the reddening covariance matrix}
\tablehead{
\colhead{$\sqrt{\lambda}$} & \colhead{$v_{u-g}$} & \colhead{$v_{g-r}$} & \colhead{$v_{r-i}$} & \colhead{$v_{i-z}$}
}
\startdata
$ 0.155 $ & $  0.88 $ & $  1.21 $ & $  0.45 $ & $  0.36 $ \\
$ 0.079 $ & $  1.23 $ & $ -0.51 $ & $ -0.41 $ & $ -0.79 $ \\
$ 0.031 $ & $  0.51 $ & $ -0.91 $ & $  0.61 $ & $  1.05 $ \\
$ 0.016 $ & $ -0.15 $ & $ -0.15 $ & $  1.35 $ & $ -0.83 $
\enddata
\tablecomments{
\label{table:cov}
Eigenvalues $\lambda$ and eigenvectors $\mathbf{v}$ for the sample covariance matrix.  Eigenvectors are normalized so that $|\mathbf{v}| = |\mathbf{R}_{\mathrm{global~fit}}|$, to facilitate comparison between $R_{a-b}$ and the first, least well constrained eigenvector, which is roughly parallel to $R_{a-b}$.
}
\end{deluxetable}

\begin{figure}[tb]
\dfplots{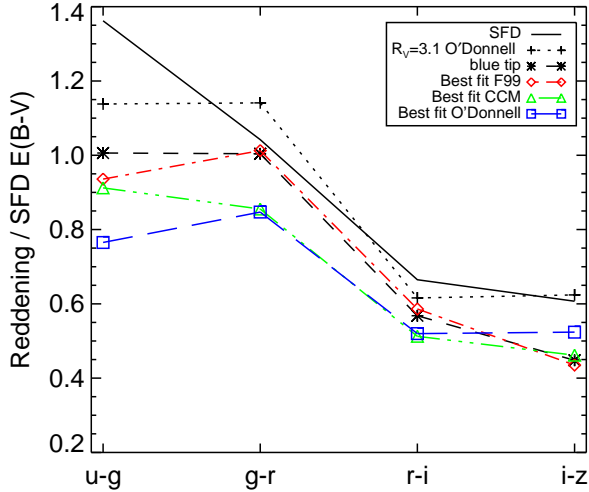}
\figcaption{
\label{fig_rvfits}
The original SFD reddening coefficients (solid line), updated coefficients according to the actual SDSS filters (crosses), and the best fit global coefficients found in this work (stars).  We fit our best fit coefficients with reddening laws according to F99 (diamonds), CCM (triangles) and O'Donnell (squares).  The F99 reddening law seems a good fit to each SDSS band (\textsection{\ref{subsubsec:rvnglobal}}).
}
\end{figure}

With this covariance matrix, the best fit reddening laws have $R_V=3.9 \pm 0.4$, $N = 0.98 \pm 0.10$ for the O'Donnell reddening law, $R_V=3.0 \pm 0.4$, $N=0.75 \pm 0.10$ for the CCM reddening law, and $R_V=3.1 \pm 0.2$, $N=0.78 \pm 0.06$ for the F99 reddening law (Fig.~\ref{fig_rvfits}).  The $\chi^2/\mathrm{dof}$ of the O'Donnell, CCM, and F99 reddening laws are 7.8, 3.5, and 2.1 respectively.  The O'Donnell and CCM reddening laws are disfavored because they predict smaller reddening differences between $r-i$ and $i-z$ relative to the data and the F99 reddening law.  The F99 reddening law fits all of the points reasonably well.

The O'Donnell reddening law in particular and CCM reddening law to a lesser extent provide poor fits to the blue tip data, and the best-fit parameters are seriously driven by the poorly-matching $r-i$ and $i-z$ data.  Accordingly, the derived $R_V$ and $N$ for these reddening laws are unphysical.

All of the fits give $R_{u-g}$ substantially less than the value given in SFD (Fig.~\ref{fig_rvfits}).  The values given in the SFD appendix were based on preliminary estimates of the system response of the SDSS, which varied somewhat from the actual system response.  Using the SDSS system response from \citet{Gunn:1998} for the $R_V = 3.1$ O'Donnell reddening law gives values different from those in SFD by 20\% in $u-g$ and 10\% in $g-r$ and $i-z$.

\begin{figure}[tb]
\dfplots{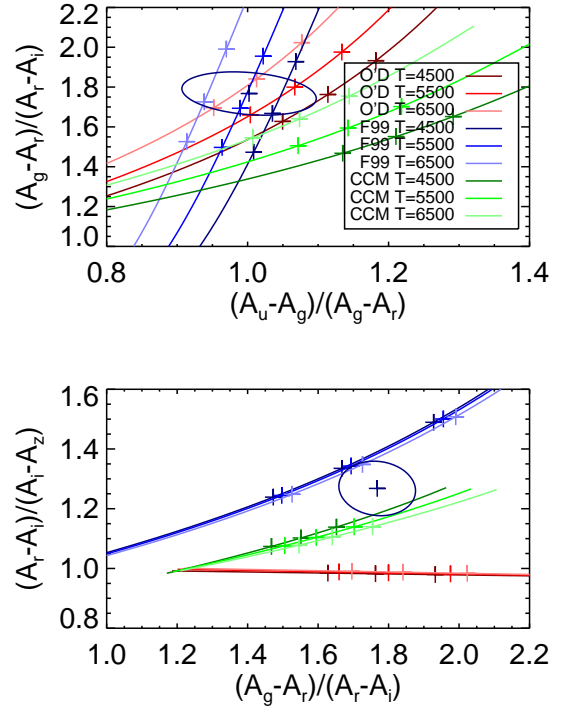}
\figcaption{
\label{fig_redlaw}
O'Donnell, CCM, and F99 reddening laws for different stars of different temperatures.  $R_V = 2.6$, $3.1$, and $3.6$ are marked with crosses.  $R_V$ increases to the left.  Ellipses are for the best fit blue tip values, with uncertainties computed from the covariance matrix of Table~\ref{table:cov}.
}
\end{figure}

The predicted $R_{a-b}$ depend somewhat on the source spectrum used (Fig.~\ref{fig_redlaw}).  We illustrate the effect of changing the source spectrum by plotting the derived reddening laws for the O'Donnell, CCM, and F99 reddening laws for Kurucz models \citep{Kurucz:1993} with $T = 6500$, $5500$, and $4500$ from the stellar spectra grid of \citet{Munari:2005}.  The blue tip best fit reddening law is relatively insensitive to $T$ over this range.

\begin{figure}[tb]
\dfplots{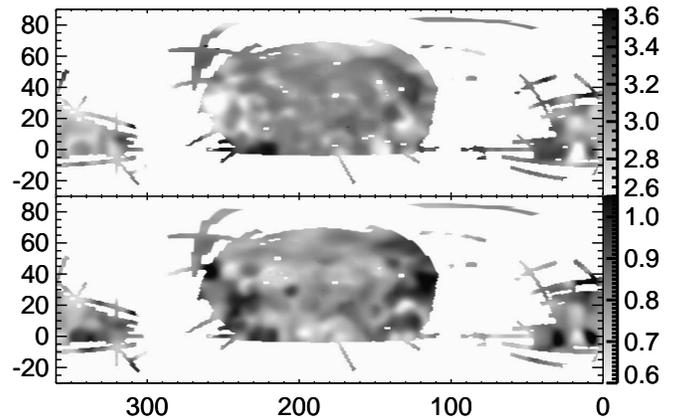}
\figcaption{
\label{fig_rvmap}
Maps of $R_V$ (top) and dust map normalization (bottom) over the SDSS footprint.  The map has been combined with a prior (\textsection{\ref{subsubsec:rvnglobal}}) to reduce the scatter in regions of low signal-to-noise.
}
\end{figure}

The maps of $R_{a-b}$ over the sky can be combined to make maps of $R_V$ and reddening law normalization over the sky (Fig.~\ref{fig_rvmap}).  To suppress noise in the measurement, we have combined the measured values with a prior of $R_V = 3.1 \pm 0.2$ and $N = 0.78 \pm 0.06$ to reduce the scatter in regions of low signal-to-noise.
One feature of the maps of $R_V$ and $N$ is that the two maps are substantially uncorrelated.  This again confirms the expectation that the ratio of $\tau_{1\mum}/\tau_{100\mum}$ does not depend on $R_V$.

\subsection{Comparison with SFD Normalization}

We can directly compare our predicted $E(B-V)$ with $\EBVSFD$ by using our best fit reddening law and accounting for the difference in source spectrum between the galaxies that SFD analyzed and the MSTO stars we analyze.  We get $E(B-V)_\mathrm{blue~tip} = 0.86 \cdot E(B-V)_{\mathrm{SFD}}$, suggesting SFD overpredicts reddening by about 14\%.  The estimated normalization uncertainty in SFD was 4\%, while we claim 8\% uncertainty in our normalization.  However, the estimated fractional uncertainty of SFD was 16\%, and we are now in a position to attribute most of this uncertainty to varying best-fit normalization.  Accordingly, it is unsurprising that the SFD normalization differs from ours by 14\%, particularly given that the footprint we analyze is different from that used in SFD, and that we have found north/south normalization differences of about 15\%.

\subsection{The SFD Temperature Correction}

The final $\EBVSFD$ takes the form $\EBVSFD = p\mathbf{I}_{corr}\mathbf{X}$, where $p$ represents a normalization coefficient, $\mathbf{I}_{corr}$ represents the destriped, zodiacal-light subtracted \IRAS~100$\mum$ flux and $\mathbf{X}$ represents a temperature correction factor.  An error in determining the temperature correction factor will lead to dust with a different best fit normalization, but with the same reddening spectrum, very much like what we see in the blue tip maps.

Accordingly, the blue tip fit residuals can plausibly be attributed to errors in the SFD temperature correction factor.  In \textsection \ref{subsec:regions}, we attempted to test the temperature correction by verifying that the best fit dust normalization was consistent for regions of different temperature according to SFD.  However, this procedure assumes that the SFD temperature correction adequately distinguishes hot and cold dust.  At high latitudes, the signal to noise of the DIRBE 100$\mum$ and especially the 240$\mum$ map are too low to construct a $100\mum/240\mum$ ratio map without substantial filtering.  The SFD ratio map was constructed by first smoothing the 100$\mum$ and 240$\mum$ maps to 1$\degree$ and then further weighting low $S/N$ pixels to a high $|b|$ average ratio.  Insofar as this procedure mixes dust of different temperatures together to a single reported SFD average temperature, it makes it difficult for us to test the SFD temperature correction using the SFD temperature map in this way.

\begin{figure}[!tb]
\dfplots{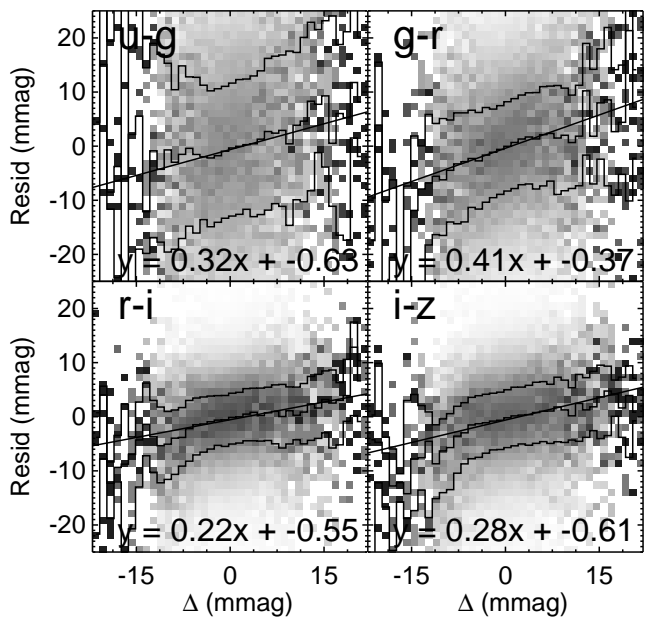}
\figcaption{
\label{fig_tempresid}
Blue tip color residuals versus change in filtered dust map $E(B-V)$ from using the FIRAS temperature correction rather than the SFD temperature correction (both in$\MMAG$).  The linear trend suggests that the temperature correction may be at fault.
}
\end{figure}

One way to test the accuracy of the temperature correction at high $|b|$ is to compare the SFD temperature correction with the temperature correction used for the FIRAS dust fits \citep{Finkbeiner:1999}.  The SFD temperature correction is weighted to constants in the north and south in regions of low signal to noise, while the FIRAS temperature correction is weighted to a $7\degree$ smoothed map in these regions.  In high signal-to-noise regions the two corrections agree.  Letting $\Xsfd$ be the SFD temperature correction and $\Xfiras$ be the FIRAS temperature correction, we can plot blue tip color residuals versus the change in predicted dust column density from switching from the SFD to the FIRAS temperature correction: $\Delta = \EBVSFD(\Xfiras/\Xsfd-1)$ (Fig.~\ref{fig_tempresid}).

Were the FIRAS temperature correction a much better predictor of dust temperature than SFD, we would hope for a linear trend between dust correction and blue tip residual, with slope equal to the best fit dust coefficients.  Instead we find slopes of approximately half the best fit coefficients, suggesting that the true temperature map is between the SFD and F99 temperature maps.  The specific values of the coefficients of the linear fits shown are not reliable because we have not accounted for the uncertainties in $\Delta$ in performing the fit, and so the slopes may be underestimated.  Regardless, the fact that a clear linear trend with significant slope exists provides strong evidence that the temperature correction at high $|b|$ is unreliable.

\begin{figure}[tb]
\dfplots{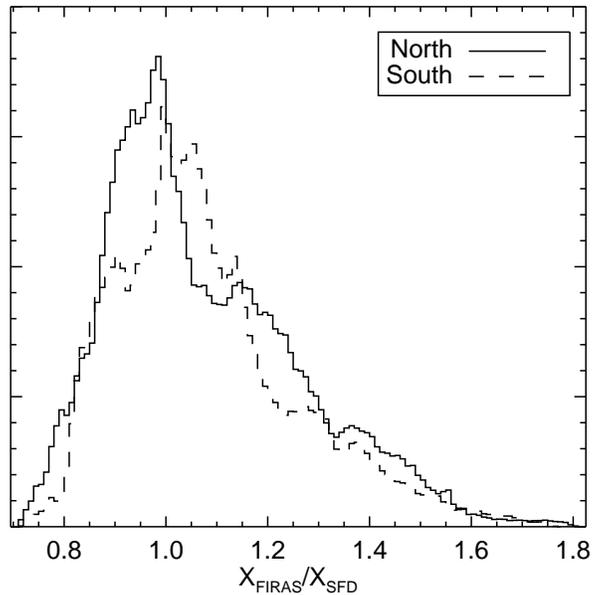}
\figcaption{
\label{fig_northsouthtemp}
The ratio $\Xfiras/\Xsfd$ in the north (solid) and in the south (dashed).  The temperature correction is systematically higher in the south than in the north by $\sim 10$\%, about compatible with the difference in best fit normalization between the north and south found in \textsection \ref{subsec:regions}.
}
\end{figure}

The FIRAS temperature correction may also explain the $\sim 15\%$ normalization difference in best fit dust extinction spectrum observed between the north and the south in \textsection \ref{subsec:regions}.  The ratio $\Xfiras/\Xsfd$ in regions with $E(B-V) < 0.05$ is about 10\% higher in the south than in the north (Fig.~\ref{fig_northsouthtemp}).

\subsection{The Dereddened Blue Tip in the North and South}

The color of the blue tip changes because of both dust and because of changing stellar populations.  If we deredden the blue tip according to SFD and the best fit $R_{a-b}$ of this work, we expect the remaining variation in blue tip color to be due to changing stellar populations (Fig.~\ref{fig_bluemaps-resid}).  

A surprising feature of the dereddened blue tip maps is that the SGC appears redder than the NGC in $g-r$, $r-i$, and $i-z$, especially outside a region in the southeast of the SGC (Table~\ref{table:nsdiff}).  This may be a genuine structure in the stellar population in the south, but we are unable to distinguish that possibility from calibration errors or errors in the dust map.  However, because the south tends to prefer a smaller SFD normalization than the north, the dereddened south would have been expected to be too blue rather than too red (Fig.~\ref{fig_dustskysubset}).  Moreover, the shape of the extra reddening in the different SDSS bands does not look like any plausible reddening law.

\begin{deluxetable}{ccccc}
\tablewidth{0pc}
\tablecaption{North/South color asymmetry}
\tablehead{
\colhead{Region} & \colhead{$u-g$} & \colhead{$g-r$} & \colhead{$r-i$} & \colhead{$i-z$}
}
\startdata
North (mag) & 0.817 & 0.228 & 0.064 & -0.033 \\
South (mag) & 0.824 & 0.250 & 0.071 & -0.020 \\
Difference (mmag) &  7.6 & 21.8 &  7.2 & 12.4

\enddata
\tablecomments{
\label{table:nsdiff}
Median blue tip color in the north and south, with $40 < |b| < 70$.  The south is systematically redder than the north.  This may be a sign of an interesting stellar structure in the south, or a calibration problem or SFD error.
}
\end{deluxetable}

\section{Conclusion}
\label{sec:conclusion}

The blue tip of the stellar locus provides a viable color standard for testing reddening.  It is also a sensitive probe of systematics in survey photometry: we detect striping artifacts from the SDSS camera columns as well as occasional runs with bad zero points.  Reddening measurements based on the stellar locus are limited because the stellar locus varies with position on the sky, due to changing stellar populations.  However, we find that we are able to overcome this limitation by looking for small-scale fluctuations that correlate with those in the SFD dust map.  By removing the best fit quadratic from the blue tip colors in each SDSS run, both survey systematics and sky-varying stellar populations seem adequately accounted for: Gaussian fits to the blue tip residuals have $\sigma$ of  18.1, 12.3, 6.9, and 7.8\MMAG, compared with the median formal statistical uncertainties $\sigma$ of 17.5, 12.5, 5.7, and 5.8$\MMAG$ in $u-g$, $g-r$, $r-i$, and $i-z$, respectively.  Because our errors are not Gaussian, in terms of $\chi^2/\mathrm{dof}$ we do worse; the fits have $\chi^2/\mathrm{dof}$ of 1.49, 1.30, 2.16, and 2.28 in the four colors.

Using these reddening measurements over the SDSS footprint, including, especially, the new, dustier, southern data, we can sensitively constrain the SFD dust map normalization in each SDSS color.  The original SFD values for $A_b/E(B-V)$ were in error because the filter responses used in the SFD appendix did not match the eventual filter responses used in the SDSS.  After taking this into account, our measurements and an $R_V=3.1$ O'Donnell reddening law match closely, except in $i-z$, where the O'Donnell reddening law overpredicts reddening dramatically.  We find an F99 reddening law provides a good fit to the data, with $R_V=3.1$ and $N=0.78$.  We recommend the use of this reddening law and normalization for use with SFD and dereddening optical data, or, for SDSS data, the use of the constants $R_{a-b}$ presented in Table~\ref{table:rab}.

This result largely vindicates the SFD normalization at high Galactic latitudes and low column densities, which had been called into question by earlier studies in dustier regions.  Except in $u-g$, where the SDSS system response assumed by SFD was in error, and in $i-z$, where the O'Donnell reddening law overpredicts reddening, we do not find that SFD overpredicts extinction by 40\%.  In $g-r$, the closest band to the SFD $B-V$ calibration, our normalization agrees with the SFD normalization within 4\%.  Extrapolating from the SDSS bands to $B-V$, we find that SFD overpredicts $E(B-V)$ by 14\%.  However, because the best fit normalization varies over the sky, this result depends on the footprint that we have analyzed, and will be different in other areas.  The fit results indicate that the normalization varies between clouds by about 10\%.

We have also made maps of the $R_V$ and $N$ over the SDSS footprint.  However, we do not yet recommend the use of these maps except in regions where the signal-to-noise is high, where we use these values to compute the $R_{a-b}$ covariance matrix.  We are actively investigating incorporating similar maps of $R_V$ and $N$ into future dust extinction maps, properly combining them with the available FIR data.

The fact that the best fit normalization of the dust map varies over the sky points to problems with the SFD temperature correction.  However, the dust extinction spectrum is relatively stable, indicating that at least in regions with $E(B-V) \lesssim 0.5$, objects can be dereddened in the optical assuming a universal extinction law to within a few percent accuracy.

The blue tip colors, dereddened according to the SFD dust map with the coefficients from this work, show no coherent residuals greater than about 30, 30, 10, and 10$\MMAG$ in the footprint we analyze, which covers most of the high-latitude sky.  Over much of the sky, residuals are within the statistical uncertainties.

These reddening measurements permit detailed tests of dust maps over a large sky area and over a range of dust temperatures and column densities.  We have been able to find clear signs of shortcomings in the SFD temperature correction.  In future work, we plan to construct new dust maps based on dust emission, constrained to best fit these measurements.  Future surveys like Pan-STARRS \citep{Kaiser:2002}, the Dark Energy Survey \citep{Flaugher:2005}, and the Large Synoptic Survey Telescope \citep{Tyson:2002} will permit this technique to be used in other bands and to fainter magnitudes with better accuracy, allowing extended wavelength coverage and admitting dustier regions near the Galactic plane to be studied.

D.F. and E.S. acknowledge support of NASA grant NNX10AD69G for this research.  M.J. acknowledges support by NASA through Hubble Fellowship grant \#HF-51255.01-A awarded by the Space Telescope Science Institute, which is operated by the Association of Universities for Research in Astronomy, Inc., for NASA, under contract NAS 5-26555.  \v{Z}.I. acknowledges support by NSF grant AST 07-07901, and \v{Z}.I. and R.G. acknowledge support by NSF grant AST 05-51161 to LSST for design and development activity.

Funding for SDSS-III has been provided by the Alfred P. Sloan Foundation, the Participating Institutions, the National Science Foundation, and the U.S. Department of Energy. The SDSS-III web site is http://www.sdss3.org/.

SDSS-III is managed by the Astrophysical Research Consortium for the Participating Institutions of the SDSS-III Collaboration including the University of Arizona, the Brazilian Participation Group, Brookhaven National Laboratory, University of Cambridge, University of Florida, the French Participation Group, the German Participation Group, the Instituto de Astrofisica de Canarias, the Michigan State/Notre Dame/JINA Participation Group, Johns Hopkins University, Lawrence Berkeley National Laboratory, Max Planck Institute for Astrophysics, New Mexico State University, New York University, the Ohio State University, University of Portsmouth, Princeton University, University of Tokyo, the University of Utah, Vanderbilt University, University of Virginia, University of Washington, and Yale University.

\appendix
\section{Using SFD to Predict Extinction in Specific Clouds}

Frequently it is useful to estimate the reddening through a particular cloud for which the reddening law is expected to be different from the Galactic average, as when the cloud has $R_V$ substantially different from three.  The correct procedure in such cases is not to use $A_V = R_V E(B-V)_\mathrm{SFD}$, because in such clouds $E(B-V) \neq E(B-V)_\mathrm{SFD}$, as SFD is ultimately a map of optical depth at 100$\mum$, and the ratio of the 100$\mum$ optical depth to $E(B-V)$ is itself a function of $R_V$.

If the cloud in question has SDSS coverage, the surest footing is to use the measured blue tip colors for that cloud to derive the extinction law in that cloud.  If SDSS coverage is not available, extrapolation from SFD or some alternative method is required.

In order to extrapolate from SFD, we have found that the best procedure is to take the SFD prediction for $E(B-V)$ and transform this to a prediction for extinction at 1$\mum$.  This gives $A_{1\mum} = 0.78 / 1.32 \cdot E(B-V)_{\mathrm{SFD}}$, where 0.78 is our best fit value for $A_{\mathrm{bluetip,~F99}}/A_{\mathrm{SFD,~O'D}}$ at $1\mum$ and 1.32 is the ratio $E(B-V)/A_{1\mum}$ for an O'Donnell $R_V = 3.1$ reddening law.  Next, reddening in any other band can be estimated by extrapolating from $A_{1\mum}$ according to an F99 reddening law.  Other reddening laws can be used, though they will require constants other than 0.78 in the above equation.

\bibliographystyle{apsrmp} 
\bibliography{blue}

\end{document}